\documentclass{emulateapj}
\usepackage{graphics}
\usepackage{subfigure}
\usepackage{amsmath}
\usepackage{bm}

\shorttitle{Particle Energy Diffusion in MHD waves}
\shortauthors{Teraki \& Asano}

\begin{document}

\title{Particle Energy Diffusion in Linear Magnetohydrodynamic Waves}

\author{Yuto Teraki\altaffilmark{1,3} and Katsuaki Asano\altaffilmark{2}}

\affil{\altaffilmark{1}National Institute of Technology, Asahikawa College, 2-2 Syunkoudai, Asahikawa, Hokkaido 071-8142, Japan; teraki@asahikawa-nct.ac.jp}
\affil{\altaffilmark{2}Institute for Cosmic Ray Research, The University of Tokyo, 5-1-5 Kashiwanoha, Kashiwa, Chiba 277-8582, Japan; asanok@icrr.u-tokyo.ac.jp}
\altaffiltext{3}{Present address: Numata Town office, 6-53, 1-3, Minami, Numata, Uryu, Hokkaido, 078-2202, Japan; y.teraki@gmail.com}

\begin{abstract}
In high-energy astronomical phenomena, the stochastic particle acceleration
by turbulences is one of the promising processes to generate non-thermal particles.
In this paper, we investigate the
energy-diffusion efficiency of relativistic particles in a temporally evolving wave ensemble
that consists of a single mode (Alfv\'en, fast or slow) of linear magnetohydrodynamic waves.
In addition to the gyroresonance with waves, the transit-time damping (TTD)
also contributes to the energy-diffusion for fast and slow-mode waves.
While the resonance condition with the TTD has been considered to be fulfilled by
a very small fraction of particles, our simulations show that
a significant fraction of particles are in the TTD resonance
owing to the resonance broadening by the mirror force, which non-resonantly
diffuses the pitch angle of particles.
When the cutoff scale in the turbulence spectrum is smaller than
the Larmor radius of a particle,
the gyroresonance is the main acceleration mechanism for all the three wave modes.
For the fast-mode, the coexistence of the gyroresonance and TTD resonance
leads to anomalous energy-diffusion.
For a particle with its Larmor radius smaller than the cutoff scale,
the gyroresonance is negligible, and the TTD becomes the dominant mechanism to diffuse its energy.
The energy-diffusion by the TTD-only resonance with fast-mode waves
agrees with the hard-sphere-like acceleration suggested in some high-energy
astronomical phenomena.

\end{abstract}

\keywords{acceleration of particles, cosmic rays, magnetohydrodynamics (MHD), turbulence}

\section{Introduction}
\label{intro}

The first-order Fermi acceleration at shock waves
is the standard model to produce non-thermal particles
in high-energy astrophysical objects.
However, several objects show a very hard photon spectrum,
which is hardly explained by the standard shock acceleration theory.
While the magnetic reconnection as an alternative model
has been discussed \citep[e.g.][]{gia08,uzd11,sir14,sir15},
the spectral-fitting models frequently suggest low magnetization in emission regions
\citep[e.g.,][]{ato96,ack16}.
Another possible model is the stochastic particle acceleration (SA) in turbulence
\citep[e.g.,][]{ach79,pt88,par95,sm98,bec06,cl06,sp08}.
The SA model has been adopted for blazars \citep{bot99,sch00,lef11,asa14,asa15,kak15,asa18},
gamma-ray bursts \citep{byk96,asa09,asat15,mur12,asa16,xu17},
pulsar wind nebula \citep{tan17}, supernova remnants \citep{liu08,fan10},
Fermi bubbles \citep{mer11,sas15},
radio galaxies \citep{or09}, and low-luminosity
active galactic nuclei (AGNs) \citep{kim14,kim15}.

Mechanisms of the SA in magnetohydrodynamic (MHD) turbulence 
have not yet been fully understood.
One ambiguous point is in the wave-particle interactions.
In the classical SA theory \citep[see][for a review]{be87}, 
the gyroresonance with the Alfv\'en waves is
considered as the main mechanism for the diffusion in the momentum space.
Non-resonant acceleration caused by the adiabatic compression of the large-scale waves
is studied by \citet{pt88} and  \citet{cl06}.
In high-energy astrophysical objects
compressional waves (fast- and slow-mode waves) can also play a role in
the particle acceleration process
\citep{cl03}.
For compressional waves, the transit-time damping \citep[TTD,][]{be58} can contribute
to accelerating particles as well.
However, the fraction of particles in the TTD resonance
has been considered very small.

The wave spectrum in turbulence, which affects the SA process,
is different in each astronomical object.
One of the most broadly 
accepted models is the Kolmogorov spectrum \citep[hereafter K41]{ko41},
$E(k)\propto k^{-5/3}$,
where $k$ is the wavenumber, and $\int E(k)dk$ is the kinetic energy of turbulence.
In the K41 model, the system is isotropic and quasi-steady,
and the cascade process is regulated by only the energy input rate
at a large scale.
However, the magnetic field
can be a crucial factor in understanding the MHD turbulence,
as the magnetic energy density is frequently comparable to the kinetic energy of turbulence.

\citet{ir63} and \citet{kr65} (hereafter IK) included the effect of the magnetic field
into their picture of isotropic turbulence.
In the IK turbulence, the Alfv\'en velocity is a parameter that characterizes the system,
and the energy distribution is expressed as $E(k)\propto k^{-3/2}$.
For fast-mode waves, the IK turbulence is a reasonable approximation
for several simulations \citep[e.g.][]{clv02}.
\citet{kl10} have shown an index $-2$ in simulations
of fast-mode turbulence.
The spectral index of fast-mode turbulence may depend on the simulation setup.
The actual index in astronomical fast-mode turbulences is highly uncertain,
while the anisotropic turbulence
so-called Goldreich--Sridhar model \citep[GS;][]{gs95} is established
for the slow and Alfv\'enic turbulence \citep[see also][]{gs97,bs05,bs06}.

\citet{yl02,yl04} argue that the gyroresonance with fast-mode waves is the 
dominant mechanism in a sub-Alfv\'enic MHD turbulence, i.e., the fluid velocity is smaller than the Alfv\'en velocity.
They estimated the diffusion coefficient in the momentum space by using the quasi linear theory (QLT),
and concluded that the contribution of gyroresonance with the fast mode is larger than other interactions 
including TTD with the fast mode.

Test particle simulations in turbulences
produced by MHD or particle-in-cell (PIC) simulations
have been carried out by many authors \citep[e.g.][]{dmi03,lq14,kim16,com18,zhd18,won19}.
For example, \citet{lq14} have claimed that the TTD resonance with slow-mode waves is the main SA mechanism
in a sub-Alfv\'enic MHD turbulence.
They pointed out the importance of the resonance broadening, due to the wave damping,
which increases the number fraction of resonating particles.
The contribution of fast waves is less than slow-mode waves, 
as the energy density of fast waves is much smaller than that of slow waves in their simulations.
However, the low-energy density of the fast mode may come from that
the turbulence forcing in their simulations is not compressive \citep{cl05}.
Note also that the Larmor radius is smaller than the grid size in the simulations in \citet{lq14}.

The arguments of \citet{yl02,yl04} and \citet{lq14} are different.
The Yan \& Lazarian papers have postulated turbulence property,
while the Lynn et al. paper generated turbulences by
MHD simulations.
Also in the test particle simulations by other authors \citep{dmi03,kim16},
each result seems different.
While PIC simulations \citep{com18,zhd18,won19} can self consistently follow
the particle motion differently from test particle simulations,
it is not easy to extract a common picture from various results.
The variety in simulated results
may be due to the difference in the physical properties such as the energy-density ratio of 
each wave mode and power-law index of the turbulence.
While MHD/PIC simulation is a powerful tool to generate turbulences,
the turbulence distribution/spectrum depends on the simulation setup.
The generated turbulences include hydrodynamical eddy motion
and multiple modes of MHD waves.
It is hard to control the final turbulence spectrum.
In addition, the interpretation of the numerical results,
especially for the resonant broadening, is not straightforward.

In this paper, we provide temporally evolving linear waves
of a single wave mode, and simulate the trajectory and energy evolution of particles.
The turbulences are described as a superposition of the Fourier modes,
which was originally proposed by \citet{gia99}.
To establish the picture of the energy-diffusion process
from a fundamental point, we neglect the nonlinear wave--wave interaction
and anisotropy of the turbulence.
This method has been widely used for the investigation of the spatial diffusion
\citep{sj11,sj15,ht15};
radiation spectra from the charged particles moving in the turbulence
\citep{ter11,ter14};
and the diffusion in the momentum space
\citep{or09,ti15}.
With this method, we can focus on each wave mode without
additional effects such as boundary effect or wave damping.
We can choose the wavelength range freely.
When we choose the shortest wavelength longer than the particle Larmor radius,
we can investigate the TTD effects without the gyroresonance.

Such large-scale turbulences may exist in high-energy astrophysical objects.
Some of those objects show violently variable non-thermal emission.
In such objects, the energy injection of turbulence may be highly variable,
and the turbulence cascade process may not develop to the small scale
comparable to the gyroradius of thermal particles.
The SA mechanism with such large-scale turbulence
is one of our subjects in this paper.
Non-resonant effects are important even when we consider resonance interactions,
as they can affect the resonance condition and it makes the mean acceleration efficiency even higher.
The non-resonant effect caused by the finite amplitude of the waves,
which are not taken into account in the QLT,
can broaden the resonance condition, and the number of the resonance particles can be higher.
For the spatial diffusions, this finite amplitude effect have been studied
\citep[e.g.][]{vol75,ac81}
as an extension of QLT.
Such effects are important not only for the incompressive (Alfv\'enic) turbulences,
but also for the compressive turbulences.
Based on \citet{vol75}, \citet{yl08}
have shown that the mirror force broadens the resonance condition of the TTD
in compressive MHD turbulence \citep[see also][]{xu18}.
This broadening mechanism is shown to be important also for the SA in this paper.

This paper is organized as follows.
In section \ref{method}, we show the calculation method.
We review
and discuss prospective property of the diffusion coefficient in section \ref{theory}.
In section \ref{result}, the numerical results are shown especially for the energy-diffusion coefficient. 
In section \ref{application}, we discuss the application of the TTD energy-diffusion
to high-energy astronomical objects.
Section \ref{discuss} provides a short discussion.

\section{Method}
\label{method}
We simulate motion of charged particles moving in an oscillating turbulence.
The field description method is based on \citet{gia99}, 
which was developed to describe a static random magnetic fluctuation.
To express oscillating MHD waves, we merge this method 
with the MHD wave decomposition method by \citet{cl03}.
We inject monoenergetic particles into the described turbulence, and solve the equation of motion.
Diffusion coefficients in the energy space is statistically obtained from the 
energy evolution of the particles interacting with the turbulence.

Our method can easily realize perturbed fields with a high resolution in time and space
with a large dynamic range.
This enables us to simulate the particle acceleration by the TTD resonance.
We cannot see the TTD resonance for test particles traveling in a fixed field obtained by
a snapshot of a MHD simulation,
while the gyroresonance can be observed in such calculations.
Moreover, in our method, we do not need a boundary condition,
which can give rise to biased effects on statistical property in particle motion.

\subsection{Field description} 
\label{field}
We describe a turbulence by a superposition of MHD waves
with a wavenumber distribution.
The MHD mode decomposition is based on the idealized MHD equations.
Assuming a wave mode, and setting an angle between the background magnetic field 
and a wave vector, we obtain the direction of the fluid velocity induced by this wave mode.
The power spectrum of the velocity fluctuation provides the amplitude of this wave mode.
Summing up the waves, we obtain the turbulence in the real space.
We explain the method below.

The background magnetic field is set to directed toward the $z$-axis as
\begin{eqnarray}
\bm{B}_0 = B_0 \hat{\bm{z}}.
\end{eqnarray}
The plasma beta for the background fluid is defined as
\begin{eqnarray}
\beta \equiv \frac{P_{\rm gas}}{B_0^2/8\pi},
\end{eqnarray}
where $P_{\rm gas} $ is the gas pressure.
The Alfv\'en velocity is written as
\begin{equation}
\mathcal{V}_{\rm A} =\frac{B_0}{\sqrt{4 \pi \rho}},
\end{equation}
where $\rho$ is the mass density of the fluid.

In this paper, we only investigate the mildly high beta region,
which is expected in high-energy 
astrophysical objects.
In the following formulae, to express the turbulence,
the parameter $\beta$ appears as only the form of the combination with the adiabatic index $\gamma_{\rm ad}$,
\begin{eqnarray}
\alpha \equiv \frac{\beta\gamma_{\rm ad}}{2} = \frac{\mathcal{V}_{\rm sd}^2}{\mathcal{V}_{\rm A}^2},
\end{eqnarray}
where $\mathcal{V}_{\rm sd}$ is the sound speed.
As a fiducial value of $\alpha$, we adopt $10$ in this paper. 
Since the adiabatic index of fully ionized plasma is ranging from 4/3 to 5/3,
$\alpha$ is roughly equal to $\beta$.

Given a wave vector $\bm{k}=k \hat{\bm{k}}$, 
we introduce the displacement vector $\hat{\bm{\xi}}$ that is the unit vector 
directed toward the fluid velocity component $\bm{V}$.
We consider the turbulence consists of a superposition of
linear waves of the MHD normal modes (Alfv\'en, fast, and slow),
neglecting the wave--wave coupling.
While the turbulence fluctuates with time, the total energy of each wave mode is conserved.

The phase velocity of the Alfv\'en wave is
\begin{equation}
\mathcal{V}_{k, {\rm A}} = \mathcal{V}_{\rm A} \cos \theta_k,
\end{equation}
where $\theta_k$ is the angle between $\bm{B_0}$ and $\bm{k}$.
The fluid velocity induced by the Alfv\'en wave is perpendicular
to both $\bm{B}_0$ and $\bm{k}$ so that
the displacement vector for the Alfv\'en wave is
\begin{eqnarray}
\label{disp-vec-A}
\hat{\bm{\xi}}_{\rm A} &=& -\hat{\bm{\phi}} \equiv
\hat{\bm{k}} \times \hat{\bm{z}}.
\end{eqnarray}

For the fast and the slow modes, the phase velocity $\mathcal{V}_{k,{\rm f/s}}$ satisfies
\begin{equation}
\mathcal{V}_{k,{\rm f/s}}^2 = \frac{1}{2}\mathcal{V}_{\rm A}^2
\left((1 + \alpha)
			      \pm \sqrt{(1 + \alpha)^2
			      - 4 \alpha \cos^2\theta_k} \right),
\end{equation}
where the upper (lower) sign in the above equation denotes the fast (slow) mode.
The fluid velocity $\bm{V}=V \hat{\bm{\xi}}$ is in the plane spanned by
$\bm{B}_0$ and $\bm{k}$.
Decomposing the velocity vector as $\bm{V}=V_z \hat{\bm{z}}+V_k \hat{\bm{k}}$,
the dispersion relation of the fast/slow wave is written as
\begin{equation}
\left(
    \begin{array}{cc}
      \omega^2 & k^2 \mathcal{V}_{\rm A}^2 \cos \theta_k  \\
      k^2 \mathcal{V}_{\rm sd}^2 \cos \theta_k & \left( \mathcal{V}_{\rm A}^2+\mathcal{V}_{\rm sd}^2 \right) k^2-\omega^2  \\
    \end{array}
  \right)
\left(
    \begin{array}{cc}
      V_z   \\
      V_k   \\
    \end{array}
  \right)=0,
\end{equation}
where $\omega$ is the wave frequency.
The velocity vector is rewritten as
$\bm{V}=(V_k +V_z \cos \theta_k) \hat{\bm{k}}-V_z \sin \theta_k \hat{{\bm{\theta}}}$,
where $\hat{{\bm{\theta}}}$ is the unit vector perpendicular to $\hat{\bm{k}}$
in the plane spanned by $\bm{B}_0$ and $\bm{k}$.
Adopting the phase velocity $\mathcal{V}_{k,{\rm f/s}}=\omega/k$, we obtain
\begin{equation}
\bm{V} \propto \hat{\bm{k}} + \frac{\sin\theta_k\cos\theta_k}
{\left(\frac{\mathcal{V}_{k,{\rm f/s}}}{\mathcal{V}_{\rm A}}\right)^2 - \cos^2\theta_k}\hat{{\bm{\theta}}}.
\end{equation}
Then, the displacement vectors for the fast and the slow mode are written as
\begin{eqnarray}
\label{disp-vec-f}
\hat{\bm{\xi}}_{\rm f} &=& \frac{1}{\sqrt{1+C_{\rm f}^2}}(\hat{\bm{k}}+ C_{\rm f} \hat{{\bm{\theta}}}), \\
\label{disp-vec-s}
\hat{\bm{\xi}}_{\rm s}&=& \frac{1}{\sqrt{1+C_{\rm s}^2}}(C_{\rm s}\hat{\bm{k}}-\hat{{\bm{\theta}}}),
\end{eqnarray}
respectively, where the factors $C_{\rm f}$ and $C_{\rm s}$ are defined as
\begin{eqnarray}
C_{\rm f} \equiv \frac{2\sin\theta_{k}\cos\theta_{k}}{2\sin^2\theta_{k}-1 +\alpha+\sqrt{(1+\alpha)^2 - 4\alpha\cos^2\theta_{k}}}, \\
C_{\rm s} \equiv -\frac{2\sin^2\theta_{k} -1 +\alpha - \sqrt{(1+\alpha)^2 - 4\alpha\cos^2\theta_{k}}}{2\sin\theta_{k}\cos\theta_{k}},
\end{eqnarray}
respectively. In the limit of $\alpha\gg1$ ($\beta \gg1$),
those values are approximated as
\begin{eqnarray}
C_{\rm f} \simeq C_{\rm s} \simeq \frac{\sin\theta_{k} \cos\theta_{k} }{\alpha}.
\end{eqnarray}
Those small factors and equations (\ref{disp-vec-f}) and  (\ref{disp-vec-s}) imply
that the fluid velocity for the fast (slow) mode is almost parallel to
$ \hat{\bm{k}}$ ($-\hat{{\bm{\theta}}}$)
under the condition of $\beta\gg1$.

The induction equation provides the perturbed magnetic field $\bm{B}=\bm{B}_0+\bm{b}$
as
$\bm{b} = (\bm{k}/\omega) \times (\bm{B}_0 \times \bm{V})$.
The magnetic field fluctuations for the Alfv\'en, fast, and slow modes are
\begin{eqnarray}
\label{ba}
\bm{b}_{\rm A}&=& \frac{V}{\mathcal{V}_{k, {\rm A}}}B_0 \cos\theta_{k} \hat{\bm{\phi}},\\
\label{bf}
\bm{b}_{\rm f}&= & -\frac{V}{\mathcal{V}_{k,{\rm f}}} B_0 \frac{C_{\rm f} \cos\theta_{k} + \sin\theta_{k}}
{\sqrt{1 + C_{\rm f}^2}} \hat{{\bm{\theta}}},  \\
\label{bs}
\bm{b}_{\rm s}&= & -\frac{V}{\mathcal{V}_{k,{\rm s}}} B_0 \frac{C_{\rm s} \sin\theta_{k} - \cos\theta_{k}}
{\sqrt{1 + C_{\rm s}^2}} \hat{{\bm{\theta}}},
\end{eqnarray}
respectively.
The divergence free condition of magnetic field $\nabla \cdot \bm{B} = 0$ is always satisfied in this description.
In our calculation method, the deviation from the condition 
$\nabla \cdot \bm{B} = 0$ does not accumulate as the numerical error,
as the field is calculated only on the particle position in each time step.

The fluid velocity $\bm{V}$ is a function of wavenumber as
\begin{eqnarray}
V^2 (k_n) = \overline{V}^2  G(k_{n})\left[\sum^N_{n=1}G(k_{n})\right]^{-1},
\end{eqnarray}
where $\overline{V}$, $G(k_{n})$ and $n$ are the spatial root mean square
of the fluid velocity,
the turbulence power function for the wave spectrum,
and the numbering of each wave, respectively.
We describe the turbulence by discrete $N$ waves,
for which we take $N=10^3$ throughout this paper.

In this paper, to concentrate on the essential effect
of turbulence excluding extra effects, we assume idealized turbulences:
isotropic or slab (only including waves of $\hat{\bm{k}}=\hat{\bm{z}}$) turbulences,
irrespectively of the wave mode.
The spectrum function is assumed as a power-law like form,
\begin{eqnarray}
G(k_{n}) = \frac{2\pi k_{n}^2\Delta k_{n}}{1 + (k_{n} L)^\nu},
\end{eqnarray}
where $L \equiv k_{\rm min}^{-1}$ and $\Delta k_{n}$ are the injection scale of the turbulence, 
and the bin width in the $k$-space, respectively.
A logarithmic spacing in $k_n$ is chosen in our computation.
For the power-law index $\nu$,
we adopt $\nu = 11/3$, $7/2$ and $4$, which correspond to the K41,
IK, and hard-sphere model, respectively.

The kinetic energy density of the fast wave is
$\rho V^2/2 \sim \rho \mathcal{V}_{\rm f}^2 (b/B_0)^2/2$.
Because $\mathcal{V}_{\rm f}/\mathcal{V}_{\rm A} \sim \sqrt{\alpha} \gg 1$,
$\rho V^2/2 \sim \alpha b^2/(8 \pi)$, which is much larger than
the turbulence magnetic energy density.
For the slow and Alfv\'en waves, the kinetic energy density is
comparable to the magnetic energy density as
$\rho V^2/2 \sim  b^2/(8 \pi)$.

From the velocity vector and magnetic field for each wave,
we can calculate the electric field from the idealized MHD condition as
\begin{eqnarray}
\bm{E} = - \frac{1}{c}\sum_{n=1}^N{\bm{V}}(k_n)
\times \left( \bm{B}_0 + \sum_{n=1}^N{\bm{b}}(k_n) \right).
\label{idealMHD}
\end{eqnarray}

To write down the electromagnetic turbulence explicitly, 
we rotate the coordinate from $(x,y,z)$ to $(x',y',z')$
as
\begin{eqnarray}
\hat{\bm{\theta}} = \hat{\bm{x}'}, \quad \hat{\bm{\phi}} = \hat{\bm{y}'},
\quad \hat{\bm{k}} = \hat{\bm{z}'},
\end{eqnarray}
\citep{gia99}.
The linear wave considered in this paper is expressed
\begin{eqnarray}
|\bm{V}(k_n)|=V(k_n) \cos\{k_n z' + \varphi(k_n) - \omega(k_n) t \},
\end{eqnarray}
where $\varphi (k_n)$ is the random phase.
The magnetic fluctuation $\bm{b}(k_n)$ also follows this form as written in equations (\ref{ba})-(\ref{bs}).

\subsection{Calculation of the diffusion coefficient}
\label{pursue}
We inject charged particles into the generated synthetic electromagnetic turbulence,
and solve the equation of motion 
\begin{eqnarray}
\frac{d\bm{p}}{dt} = e\left(\bm{E} + \frac{\bm{v}}{c}\times \bm{B}\right),
\end{eqnarray}
by the Buneman--Borris method, where $\bm{v}$ is the velocity of the particle.
The initial Lorentz factor is set to monoenergetic and
the distribution of the initial velocity direction is isotropic.
To evaluate the diffusion coefficient, we calculate the mean displacement 
of the Lorentz factor of $\left< (\Delta \gamma)^2 \right> $.
If the particle accelerated diffusively, it evolves as
\begin{eqnarray}
\left< (\Delta \gamma)^2 \right>  = 2D_{\gamma\gamma} t,
\end{eqnarray}
where $D_{\gamma\gamma}$ is the energy (Lorentz factor) diffusion coefficient.

\section{Theoretical Description}
\label{theory}

Before showing the numerical results, we review
and discuss prospective properties of the diffusion coefficient.
In a disturbed electromagnetic field, the acceleration of particles
is done by the electric field.
However, if a particle interacts with a periodic electric field
induced by plane waves, the net energy change becomes zero.
So the energy-diffusion is inefficient in such a situation.
The resonance with a wave is essential for the particle acceleration.
Given the gyro frequency of the particle $\Omega \equiv e B_0/(\gamma mc)$,
the resonance condition is written as 
\begin{equation}
\label{res-cond}
\omega - k_\parallel v_\parallel = n \Omega,
\end{equation}
where $k_\parallel$ and $v_\parallel$ are the parallel components of
the wave vector and the particle velocity
to the background magnetic field, respectively.
The integer $n$ provides the resonance conditions.

The gyroresonance corresponds to the case of $n = \pm 1, \pm 2,...$.
Usually, the acceleration is dominated by the resonance with $n = \pm1$.
When the particle is relativistic and the wave is nonrelativistic
(the phase velocity $\mathcal{V}_{\rm ph} \ll c$), 
the resonance condition with $n = \pm1$  can be approximated as $k_\parallel^{-1} \sim r_{\rm L}$,
where $r_{\rm L}$ is the Larmor radius of the particle.

On the other hand, the TTD resonance condition is $n = 0$.
The compressional waves (fast and slow waves)
can accelerate particles via the TTD resonance,
in which $v_\parallel$ equals to the parallel component of the phase velocity
$\mathcal{V}_{{\rm ph}, \parallel}$.

The power spectrum of the magnetic turbulence is defined
as
\begin{equation}
\delta B^2=\sum_{n=1}^N b^2(k_n) =\int_{k_{\rm min}}^{k_{\rm max}}dk P_B (k).
\end{equation}
When we write $P_B (k) \propto k^{-q}$, this implies $q= \nu-2$.
In this section, we adopt an approximation:
\begin{equation}
P_B (k) \simeq (q-1) \frac{\delta B^2}{k_{\rm min}} \left( \frac{k}{k_{\rm min}} \right)^{-q}.
\label{PBk}
\end{equation}

\subsection{Energy Gain}
\label{enegain}

\begin{figure}[!htb]
\centering
\epsscale{1.1}
\plotone{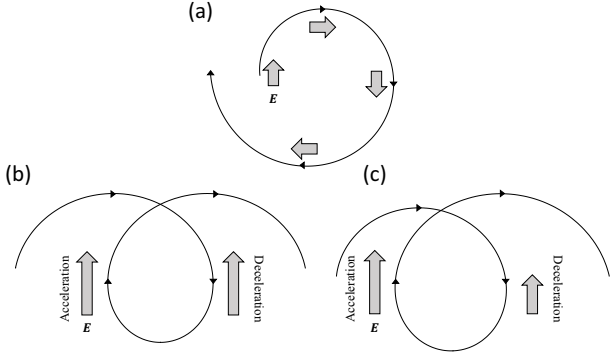}
\caption{Schematic pictures of the particle energy gain via interaction
with MHD waves. The charge of the particle is positive.
The magnetic field is perpendicular to the page,
and the particle trajectories are projected to this plane.
(a) Gyroresonance case ($\Omega \simeq \pm k_\parallel v_\parallel$).
(b) TTD case ($v_\parallel \simeq \omega/k_\parallel$)
with an Alfv\'en wave or a slow wave with $k_\perp=0$.
(c) TTD case with a fast/slow wave with $k_\perp \neq 0$.}
\label{fig-schi}
\end{figure}

Figure \ref{fig-schi} shows the energy change processes via interaction
with electric fields induced by MHD waves.
In the gyroresonant case, the electric field changes its direction
resonantly with the gyro motion.
As Figure \ref{fig-schi} (a) shows, the Larmor radius
grows according to the growth of the particle energy.

On the other hand, even if a particle is in the TTD resonance
($v_\parallel \simeq \omega/k_\parallel$),
an Alfv\'en wave and a slow
wave\footnote[4]{The electric field is zero for a fast wave with $k_\perp=0$.}
with $k_\perp = 0$ do not accelerate the particle.
In this case, the electric and magnetic fields the particle experiences
are constant.
As Figure \ref{fig-schi} (b) shows, the effects of acceleration
and deceleration cancel out in one gyro period.
In the plane perpendicular to the magnetic field ($\bm{B}_0+\bm{b}$),
we simply observe the $\bm{E}\times\bm{B}$ drift
motion.\footnote[5]{In the comoving frame with those parallel waves
along $\bm{B}_0$,
the electric field vanishes. In this frame, the TTD resonance particles
show simple gyro motion around a vector $\bm{B}_0+\bm{b}$
tilted to the background $\bm{B}_0$.
In the laboratory frame, where the electric field is finite,
the drift motion is seen in the plane perpendicular to $\bm{B}_0+\bm{b}$.
As a result, the particles propagate along $\bm{B}_0$,
though the field is tilted as $\bm{B}_0+\bm{b}$.}

To increase the particle energy, a difference of the electric field
in the perpendicular direction is required (see Figure \ref{fig-schi} (c)),
which means a finite parallel value of $\nabla \times \bm{E}$.
From the induction equation,
this condition is equivalent to a finite value of $\dot{b}_\parallel$,
which is zero in an Alfv\'en wave and parallel slow/fast wave.

The energy gain per one gyro motion is $e \oint \bm{E} \cdot d\bm{l} \sim e |\Delta \bm{E}| r_{\rm L}$.
The electric field difference is $|\Delta \bm{E}| \sim r_{\rm L} dE/dx_\perp
\sim |\bm{V}\times \bm{B}_0| k_\perp r_{\rm L}/c$.
During the TTD resonance,
the number of gyration is roughly $N_{\rm gyro}\sim(k_\parallel r_{\rm L})^{-1}$,
so that the energy change per interaction
becomes $|e\Delta \bm{E}| r_{\rm L} N_{\rm gyro} \propto V B_0 \tan{\theta_k} r_{\rm L}
\propto V \tan{\theta_k} \gamma$.
Though a lower magnetic field leads to a lower electric field,
the resultant larger Larmor radius ($\propto B_0^{-1}$)
extends the scale $\Delta x_\perp$, which enhances $|\Delta \bm{E}|$.
Therefore, the energy change is independent of $B_0$,
which agrees with the picture of the second order Fermi 
acceleration\footnote[6]{In the shock acceleration,
the motional electric field in the upstream for an observer
in the downstream frame is proportional to the magnetic field.
For a lower magnetic field, the electric field becomes lower,
but the resultant large Larmor radius extends the residence time of particles
in the upstream. Thus, the energy gain per one cycle $\Delta \gamma \propto \gamma$
is independent of the magnetic field as is well known.}
($\Delta \gamma \propto \gamma$).

However, if $k_\perp r_{\rm L} \gg 1$,
too frequent changes of the electric field sign in one gyro motion
reduce the net energy change.
To accelerate particles via the TTD resonance,
the magnetic field should be strong enough to keep $k_\perp r_{\rm L} < 1$.

\subsection{Gyroresonance}
\label{gyro-sec}

The analytical description of the energy-diffusion via gyroresonance with MHD waves
has been shown by many authors \citep[see e.g.,][]{mel74,be87,sch89,sch89b}.
We do not need to review the detailed kinematics in this process.
A particle of a Lorentz factor $\gamma$ interacts
with mainly waves of a wavenumber
\begin{equation}
k_{\rm res} \equiv \frac{e B_0}{\gamma m c^2},
\end{equation}
in gyroresonance.

From QLT, the diffusion coefficient with gyroresonance
is represented as
\begin{eqnarray}
\label{dgg-qlt-slab}
D_{\gamma\gamma} &=&  \frac{\pi}{12}\gamma^2\left(\frac{\mathcal{V}_{\rm ph}}{c}\right)^2
	\left(\frac{k_{\rm res}P_B(k_{\rm res})}{B_0^2}\right) \frac{e B_0}{\gamma m c},
\end{eqnarray}
where $\mathcal{V}_{\rm ph}$ is
the phase velocity of the scattering agent.
Hereafter, we express the parameter $k_{\rm min}$
as
\begin{eqnarray}
k_{\rm min}=\eta \frac{e B_0}{m c^2},
\end{eqnarray}
then
\begin{eqnarray}
D_{\gamma\gamma} 
\simeq \frac{\pi}{12} (q-1) \left(\frac{\mathcal{V}_{\rm ph}}{c}\right)^2
\frac{\delta B^2}{B_0^2} \eta^{q-1} \gamma^q \frac{e B_0}{m c}.
\end{eqnarray}
From equations (\ref{ba})--(\ref{bs}),
$\delta B^2/B_0^2 \sim (\overline{V}/\mathcal{V}_{\rm ph})^2$.
Then, irrespectively of the wave mode,
the diffusion coefficient is written as
\begin{eqnarray}
D_{\gamma\gamma} 
&\simeq& \frac{\pi}{12} (q-1) \left(\frac{\overline{V}}{c}\right)^2
\eta^{q-1} \gamma^q \frac{e B_0}{m c} \label{dgggyro00}\\
&\simeq& 3.76\times 10^{-7} \left(\frac{\overline{V}}{10^{-3} c}\right)^2
\left(\frac{\eta}{10^{-2}}\right)^{2/3}
\left(\frac{\gamma}{10}\right)^{5/3} \frac{e B_0}{m c},\nonumber \\
\end{eqnarray}
where we have adopted $q=5/3$.

\subsection{TTD Resonance}
\label{TTD-sec}

\subsubsection{Resonance Broadening}

The fraction of particles that fulfill the resonance condition
$v_\parallel=\mathcal{V}_{{\rm ph}, \parallel} \ll c$
is very small for the isotropic particle distribution, 
so that the TTD is usually ignored in QLT.
The condition is rewritten as $\cos\theta_{\rm p} \sim \mathcal{V}_{{\rm ph}, \parallel}/c$,
where $\theta_{\rm p}$ is the pitch angle.
However, even if the initial pitch angle is smaller
than $\cos^{-1}{(\mathcal{V}_{{\rm ph}, \parallel}/c)}$,
the finite amplitude of the turbulence can change the pitch angle,
and make the particles put into the TTD resonance.

The non-resonant pitch angle change comes from the mirror force.
We write the initial parallel and perpendicular components of the particle velocity
as $v_{\parallel, 0}$ and $v_{\perp,0}$, respectively.
According to $v_\parallel^2+v_\perp^2=c^2$ and the adiabatic invariance $p_\perp^2/B$,
the fluctuating magnetic field $B_0+\delta B$ changes the velocity as
\begin{equation}
\label{res-borad}
v_\parallel^2 = v_{\parallel, 0}^2 - v_{\perp,0}^2 \frac{\delta B}{B_0}.
\end{equation}
Denoting the square of the mean amplitude of the disturbed field component
as $\left< (\delta B)^2 \right>$, we introduce
the Bohm factor as
\begin{equation}
\xi \equiv \frac{B_0^2}{\left< (\delta B)^2 \right>}.
\end{equation}
To make particles put into the TTD resonance with non-relativistic
waves ($\mathcal{V}_{{\rm ph}, \parallel} \ll c$), $v_\parallel \simeq 0$
is required.
If the initial pitch angle is larger than $\theta_{\rm min}$,
which is defined as
\begin{equation}
\sin^2{\theta_{\rm min}} \equiv \frac{1}{1+\xi^{-1/2}},
\end{equation}
$v_\parallel \simeq 0$ is realized by the mirror force.

In the isotropic distribution,
the fraction of particles that satisfy
$\theta_{\rm min}<\theta_{\rm p}<\pi-\theta_{\rm min}$ is
\begin{equation}
f_{\rm res}=\cos{\theta_{\rm min}}=\frac{1}{\sqrt{1+\xi^{1/2}}},
\end{equation}
when $\cos{\theta_{\rm min}}> \mathcal{V}_{\rm ph}/c$.
Note that the resonance broadening does not depend on the wave scale.

\begin{figure}[!htb]
\centering
\epsscale{1.1}
\plotone{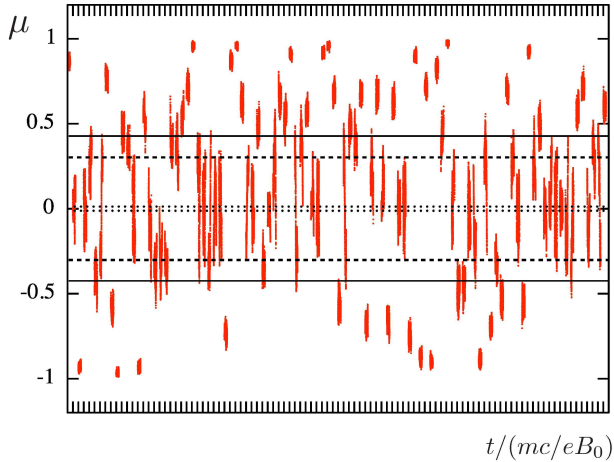}
\caption{Time evolution of the pitch angle of the each particles.
We adopt isotropic slow-mode waves,
whose parameters are $q=5/3$, $\alpha=10$, $\eta=10^{-4}$, $k_{\rm max}=100 k_{\rm min}$,
$\mathcal{V}_{\rm A}=10^{-2} c$, and $\overline{V}=10^{-3} c$.
The vertical axis is the pitch angle cosine.
We inject particles of $\gamma=10$
and follow each particle for a time interval of $10^6 mc/(e B_0)$,
which corresponds to the grid scale on the horizontal axis.
We plot the evolution of $\mu$ for randomly selected 100 particles
with different time offsets.
The horizontal dashed and solid lines correspond to $\cos{\theta_{\rm min}} \simeq \pm 0.3$
and $\sqrt{2} \cos{\theta_{\rm min}}$, respectively.
The dotted lines are $\mu = \mathcal{V}_{\rm A}/c =\pm 0.01$,
which is the TTD resonance condition without the resonance broadening.
}
\label{mu-evo}
\end{figure}

To demonstrate the resonance broadening,
we provide isotropic slow-mode waves with $\nu = 11/3$ ($q=5/3$)
between $k=k_{\rm min}$ and $k_{\rm max}=100 k_{\rm min}$,
where $\eta=10^{-4}$, and inject particles with the initial Lorentz factor $10$,
for which $k_{\rm res}=10 k_{\rm max}$.
In this case, the pitch angle diffusion due to gyroresonance can be neglected.
With the method explained in \S \ref{method}, we follow the particle trajectory.
The results are shown in Figure \ref{mu-evo}.
Here, we have assumed $\alpha=10$, $\mathcal{V}_{\rm A}=10^{-2} c$,
and $\overline{V}=0.1 \mathcal{V}_{\rm A}$.
For $\alpha \gg 1$, the phase velocity of the slow wave is roughly $\mathcal{V}_{\rm A}$,
so that the Bohm factor is approximated as $\xi \simeq (\mathcal{V}_{\rm A}/\overline{V})^2=100$
in this case.

Figure \ref{mu-evo} shows that
particles with $|\mu|=|\cos\theta_{\rm p}|$ smaller than $\cos{\theta_{\rm min}} \simeq 0.3$
(dashed line)
can penetrate into the region $|\mu| < \mathcal{V}_{\rm A}/c=0.01$ (dotted line)
owing to the mirror force.
Note that the pitch angle starting from $\theta_{\rm p}=\theta_{\rm min}$
can become as small as $\cos\theta_{\rm p}=\sqrt{2} \cos{\theta_{\rm min}}$
(obtained from equation (\ref{res-borad}) with negative $\delta B$).
Although $\xi =100$  is significantly large,
the effect of the mirror force efficiently enlarges the resonance particle number.
The example in Figure \ref{mu-evo} is for the slow mode.
The results for the fast mode are also similar to the slow-mode results.

Let us estimate the diffusion coefficient of the pitch angle in this case.
The combination of the adiabatic constant $p_\perp^2/B$ and
the energy $p_\perp^2+p_\parallel^2+m^2 c^2$
provides
\begin{eqnarray}
\frac{\Delta p}{\delta t} \sim \frac{p_{\perp}v_\perp}{2B}\nabla_\parallel |B_\parallel| \sim pv_\perp k_\parallel \frac{\delta B(k)}{B_0},
\end{eqnarray}
where $\delta B(k) \equiv \sqrt{k P_{\rm B}(k)}$.
During the transit-time of the wave,
\begin{eqnarray}
\delta t \sim \frac{1}{v_\parallel k_\parallel},
\end{eqnarray}
the pitch angle changes as
\begin{eqnarray}
\Delta \mu \sim  \frac{\Delta p}{p} \sim \frac{v_\perp}{v_\parallel}  \frac{\delta B(k)}{B_0},
\end{eqnarray}
so that the resonance broadening is dominated by the largest scale for $q>1$.
The pitch angle diffusion coefficient is written as 
\begin{equation}
\label{dmm}
D_{\mu\mu} \sim \sum_n \frac{(\Delta \mu)^2}{\delta t}\sim \sum_n v_\perp^2k_\parallel^2\left(\frac{k_nP_B(k_n)}{B_0^2}\right)\cdot \delta t.
\end{equation}

\subsubsection{Energy Diffusion Coefficient}
\label{theoryenedif}

\begin{figure}[!htb]
\centering
\epsscale{1.1}
\plotone{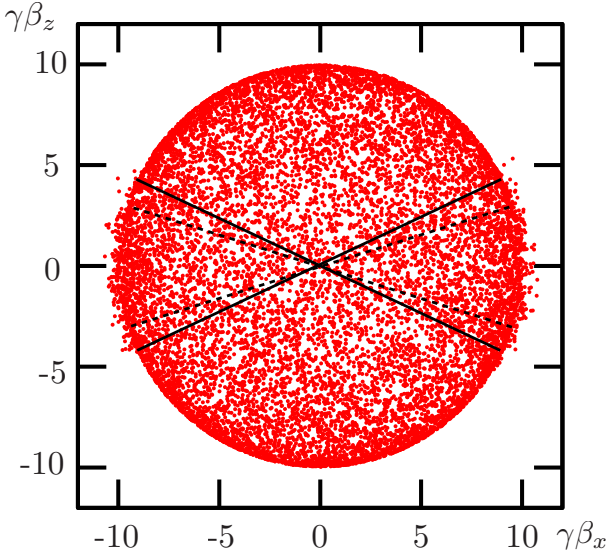}
\caption{Distribution of $\gamma \bm{\beta}$
of 12,800 particles after elapsed time of $10^6 mc/eB_0$
in the pure slow-mode waves.
The simulation is the same as that in Fig. \ref{mu-evo}.
Horizontal axis is $\gamma \beta_x$, and vertical axis is $\gamma \beta_z$.
The background magnetic field is parallel to $\bm{z}$.
The long dashed lines and solid lines correspond
to $\cos{\theta_{\rm min}}$ and $\sqrt{2} \cos{\theta_{\rm min}}$,
the critera of the resonance broadening.
}
\label{gam-beta-s}
\end{figure}

Figure \ref{gam-beta-s} shows the particle distribution
in a cross section of the momentum space for the same simulation
as that in Figure \ref{mu-evo}, where slow waves are isotropically propagating.
As discussed in the previous subsection, particles within $|\mu| < \cos{\theta_{\rm min}}$
can resonate with the waves.
We can see that only such resonating particles are exuded outside the equi-energy circle of
$\gamma \sqrt{\beta_x^2 + \beta_z^2} = 10$, which is the initial energy
at injection. In this subsection,
we describe the energy-diffusion coefficient in the TTD resonance.

\begin{figure}[!htb]
\centering
\epsscale{1.1}
\plotone{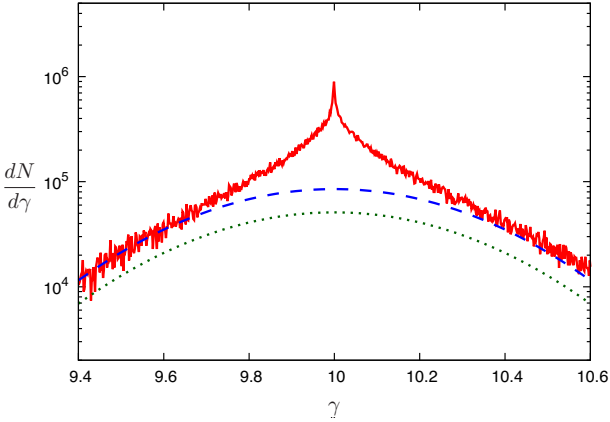}
\caption{Energy distribution of particles at $t = 10^6 m c/(e B_0)$
in the pure fast-mode waves.
The parameters are $q=3/2$, $\alpha=10$, $\eta=10^{-4}$, $k_{\rm max}=100 k_{\rm min}$,
$\mathcal{V}_{\rm A}=10^{-2} c$, and $\overline{V}=10^{-3} c$.
The initial Lorentz factor is 10.
The blue dashed line and green dotted lines are the Gaussian curves with dispersion $\sigma = 0.5$,
and particle numbers $0.5N_{\rm all}$ and $0.3N_{\rm all}$, respectively
(the total particle number $N_{\rm all} = 12,800$).
}
\label{bipole}
\end{figure}

Figure \ref{bipole} shows another demonstration of the resonance broadening
for fast waves.
Also in this case, the gyroresonance is negligible effect ($k_{\rm res}=10 k_{\rm max}$).
Though the phase velocity is faster than the phase velocity for slow wave,
$\mathcal{V}_{\rm ph}/c \sim \sqrt{\alpha} \mathcal{V}_{\rm A}/c \simeq 0.03$
is still lower than $\cos{\theta_{\rm min}}\simeq 0.3$.
As the figure shows, roughly 50\% of particles are in the diffusion process,
and rest of particles remains around the initial energy.
The resonance fraction is slightly larger than the simple estimate
$\cos{\theta_{\rm min}}$ or $\sqrt{2}\cos{\theta_{\rm min}}$.

Because the energy-diffusion due to the TTD resonance is consistent with the second order
Fermi acceleration,
we can write the energy change in the timescale $\delta t \sim (v_\parallel k_\parallel)^{-1}$
by considering the Lorentz transformation 
between the wave rest frame and Laboratory frame before and
after the pitch angle change. The change of the angle $\Delta \mu$
in the wave rest frame yields the energy change as
\begin{eqnarray}
\Delta \gamma \sim \frac{\mathcal{V}_{\rm ph}}{c}\gamma\Delta \mu.
\end{eqnarray}
The fraction of particle that can resonate with waves can be written as
 \begin{eqnarray}
f_{\rm res}=\max\left( \frac{\mathcal{V}_{\rm ph}}{c},\frac{1}{\sqrt{1+\xi^{1/2}}} \right).
\end{eqnarray}
Then, the energy-diffusion coefficient is estimated as
 \begin{eqnarray}
 D_{\gamma\gamma} \sim \sum_n \frac{(\Delta \gamma)^2}{\delta t} f_{\rm res} \sim
\sum_n \gamma^2\left(\frac{\mathcal{V}_{\rm ph}}{c}\right)^2 \frac{(\Delta \mu)^2}{\delta t} f_{\rm res}.
\end{eqnarray}
From equation (\ref{dmm}), we obtain
 \begin{eqnarray}
 D_{\gamma\gamma} \sim
\sum_n \gamma^2\left(\frac{\mathcal{V}_{\rm ph}}{c}\right)^2 c k_n \left(\frac{k_nP_B(k_n)}{B_0^2}\right) f_{\rm res}.
\label{dgg0}
\end{eqnarray}

Each term in equation (\ref{dgg0}) is proportional to $ k_n^{2-q}$,
so that the largest $k_n$ contributes dominantly for $q<2$.
As we mentioned in the previous subsection,
the wave power absorption by particles is strongly suppressed for $k_\perp r_{\rm L}>1$.
Therefore, for $k_{\rm res}<k_{\rm max}$,
the diffusion coefficient is approximated as
\begin{equation}
\label{dgg-res}
 D_{\gamma\gamma} \sim \gamma^2\left(\frac{\mathcal{V}_{\rm ph}}{c}\right)^2 c k_{\rm res} \left(\frac{k_{\rm res}P_B(k_{\rm res})}{B_0^2}\right)
f_{\rm res},
\end{equation}
which is almost the same as the coefficient in the gyroresonance case,
equation (\ref{dgg-qlt-slab}), except for the factor $f_{\rm res}$.
Finally the expression becomes a similar equation to equation (\ref{dgggyro00}) as
\begin{eqnarray}
D_{\gamma\gamma} 
&\sim& f_{\rm res} (q-1) \left(\frac{\overline{V}}{c}\right)^2
\eta^{q-1} \gamma^q \frac{e B_0}{m c}.
\end{eqnarray}

On the other hand, for $k_{\rm res}>k_{\rm max}$, we obtain
\begin{equation}
\label{dgg-large}
 D_{\gamma\gamma} \sim \gamma^2\left(\frac{\mathcal{V}_{\rm ph}}{c}\right)^2 c k_{\rm max} \left(\frac{k_{\rm max}P_B(k_{\rm max})}{B_0^2}\right)
f_{\rm res}. \\
 \end{equation}
From equation (\ref{PBk}), this is rewritten as
\begin{equation}
 D_{\gamma\gamma} \sim f_{\rm res} (q-1) \gamma^2 \left(\frac{\overline{V}}{c}\right)^2 c k_{\rm max}  
 	\left(\frac{k_{\rm max}}{k_{\rm min}}\right)^{1-q}.
\end{equation}

The precise numerical factor for $D_{\gamma\gamma}$ may depend on
the wave mode. We have derived the above rough formulation
aside the concept of the resonance discussed in section \ref{enegain}.
However, as we have discussed,
a finite value of $k_\perp$ is required to change the particle energy
via the TTD resonance.
In the limit of $\alpha \gg 1$,
the direction of fluid motion is almost parallel (perpendicular) to
the wave vector $\bm{k}$ for the fast (slow) mode.
For a significantly large $k_\perp$,
the fast mode is advantageous to induce a larger electric field
than the slow mode. We expect a larger energy-diffusion by the TTD resonance
in the fast mode than the slow mode.

\section{Diffusion Coefficients in Simulations}
\label{result}
In this section, we show the energy-diffusion process
obtained from our numerical simulations with the method explained in section \ref{method}.
The injected turbulence energy at a large scale is transferred to small scale turbulences.
This cascade process may be interrupted at a certain scale
by energy dissipation processes like the TTD.
Below this scale, the power spectrum becomes steep.
This scale practically
corresponds to the maximum wavenumber $k_{\rm max}$.

Hereafter, given an energy of particles,
we consider two cases for the maximum wavenumber $k_{\rm max}$:
one with gyroresonance
($k_{\rm res}<k_{\rm max}$) and one without gyroresonance ($k_{\rm res}>k_{\rm max}$).

\subsection{With Gyroresonance}
\label{small}

\begin{figure}[!h]
\centering
\epsscale{1.0}
\plotone{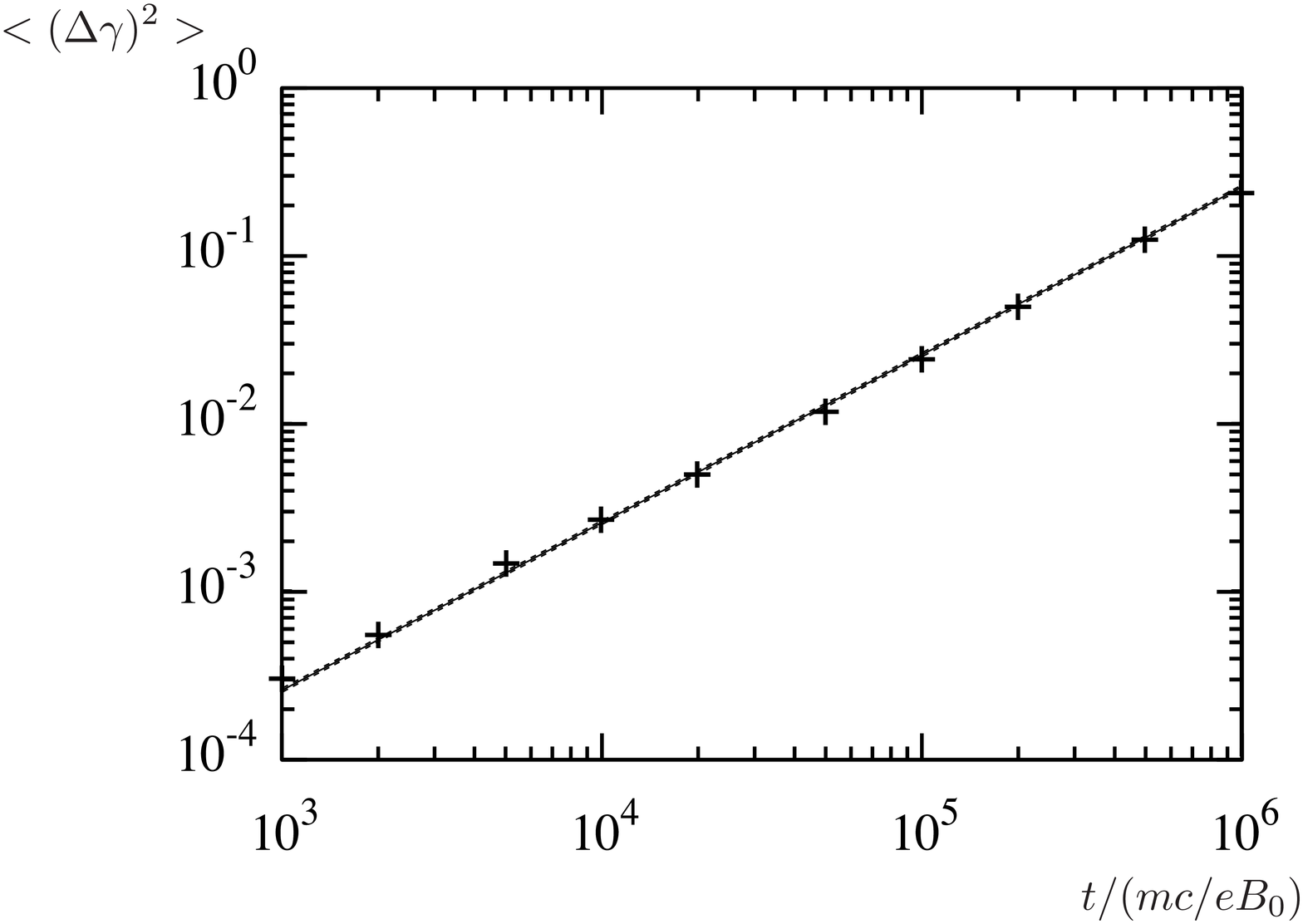}
\plotone{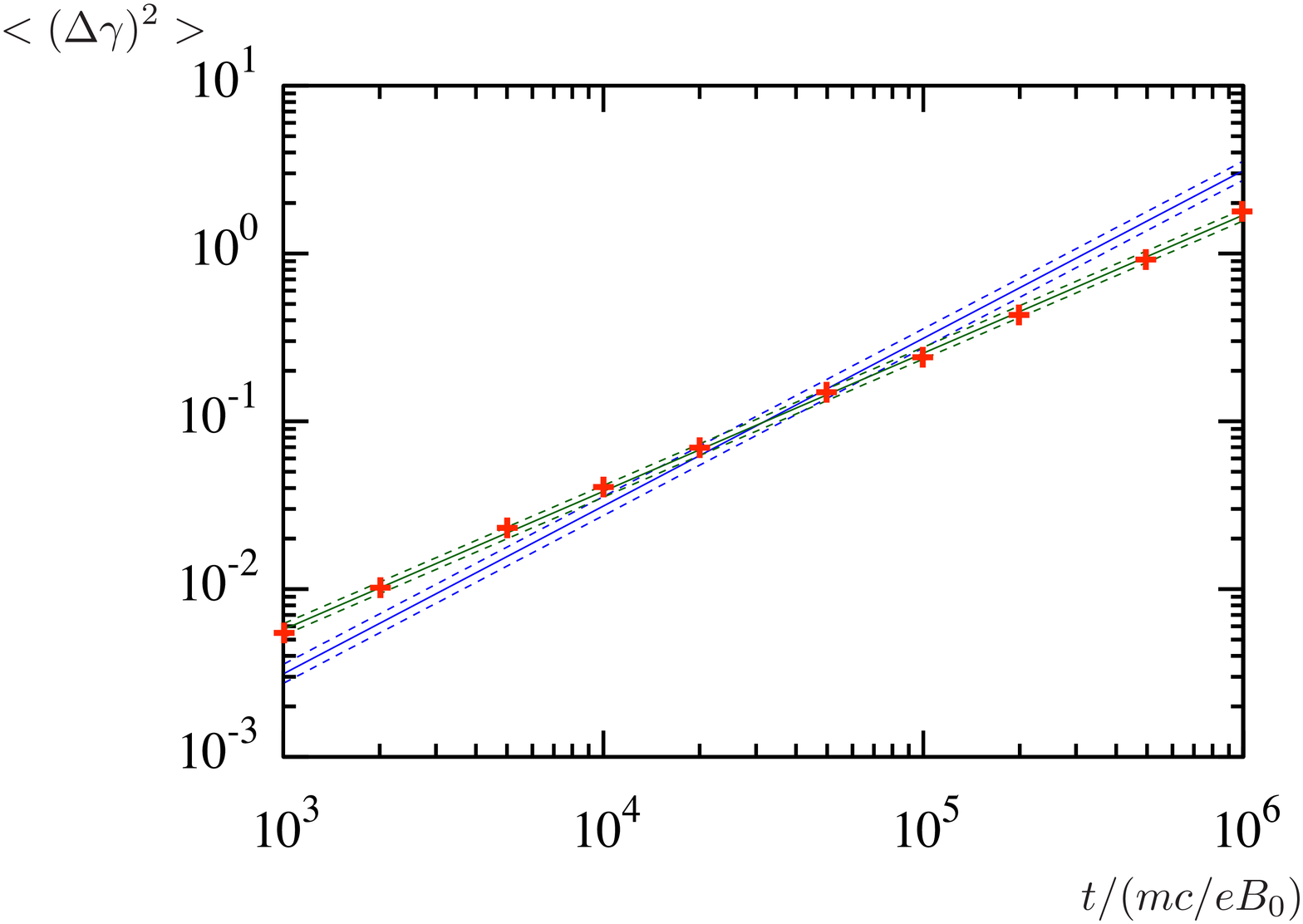}
\plotone{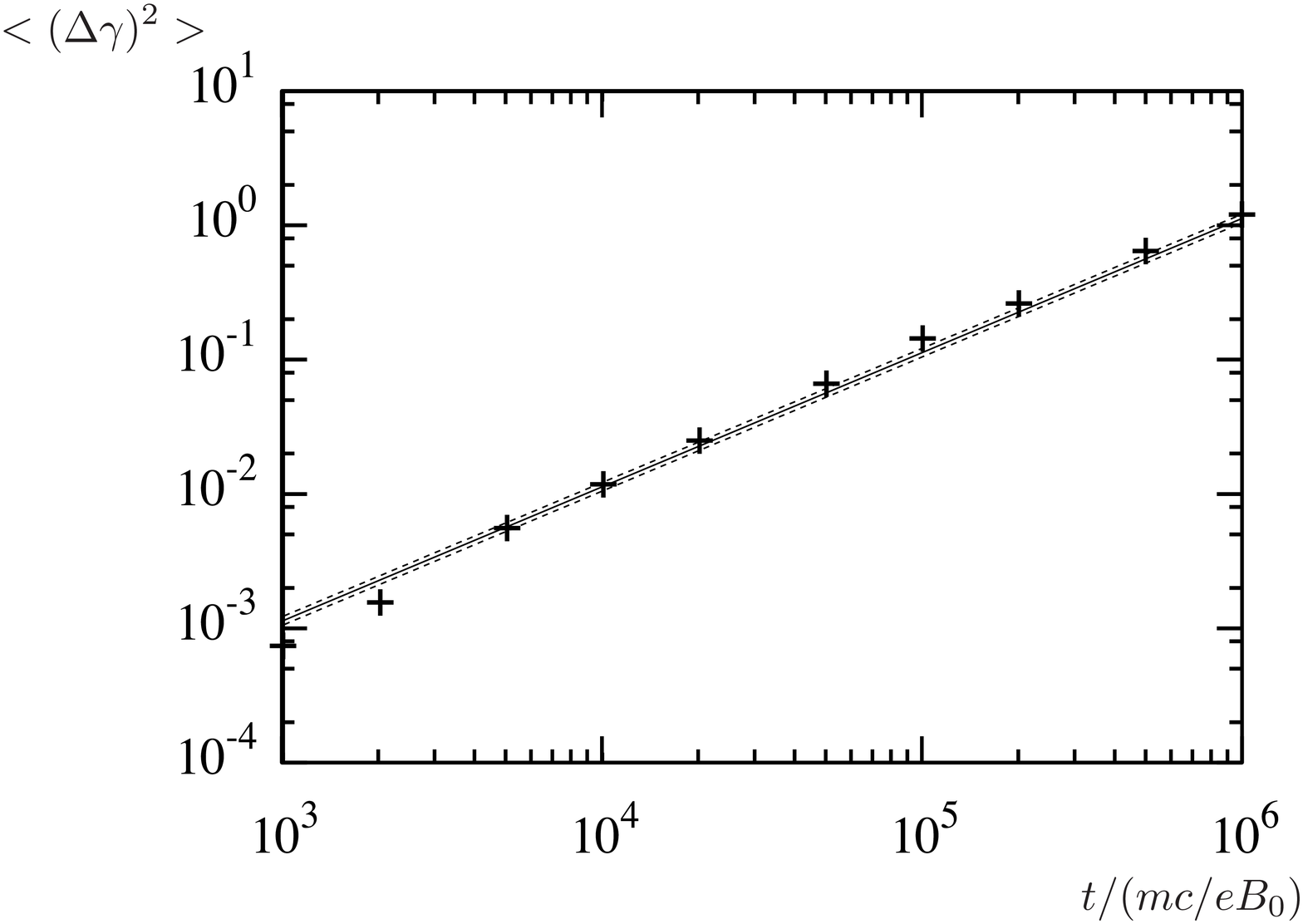}
\caption{Evolutions of the dispersion of Lorentz factor
$\gamma$ in turbulences with $q=5/3$, $\alpha=10$, $\eta=10^{-2}$,
$k_{\rm max}=100 k_{\rm min}$,
$\mathcal{V}_{\rm A}=10^{-2} c$, and $\overline{V}=10^{-3} c$.
The initial Lorentz factor is 10 ($k_{\rm res}=0.1 k_{\rm max}$).
The upper, middle and lower panels show the results for
pure Alfv\'en, fast- and slow-mode waves, respectively.
The solid and dashed lines indicate the fitting results.
For the Alfv\'en and the slow modes, the numerical results agree
with the standard diffusion behavior $\left< (\Delta \gamma)^2 \right> \propto t$.
For the fast mode (middle), while the blue lines of
$\left< (\Delta \gamma)^2 \right> \propto t$ are deviated from the numerical results,
the green lines of $\left< (\Delta \gamma)^2 \right> \propto t^{0.82}$
well follow the results.
}
\label{dgg-evol}
\end{figure}

For particles of $k_{\rm res}<k_{\rm max}$, both the gyroresonance and
TTD resonance contribute to the particle energy-diffusion.
As shown in Figure \ref{dgg-evol}, we follow the motion of 
$N=1,280$ particles injected with $\gamma=10$
and obtain the temporal evolution
of the dispersion of $\gamma$ for each run.
Between $t =10^3 m c/(e B_0)$ and $10^6 m c/(e B_0)$, we fit the evolution of the dispersion
with $\left< (\Delta \gamma)^2 \right> \propto t$
and estimate the energy-diffusion coefficient.

As Figure \ref{dgg-evol} shows, the dispersion
for the Alfv\'en mode, where only the gyroresonance
contributes to the energy-diffusion,
precisely follow the analytical evolution as $\left< (\Delta \gamma)^2 \right>
\propto t$.
We have not found signature of the advection in the energy space
like the first-order Fermi acceleration, which leads to $\left< (\Delta \gamma)^2 \right>
\propto t^2$.
However, the evolution for the fast mode significantly
deviates from the standard diffusion behavior.
The dispersion in the fast mode rather evolves as
$\left< (\Delta \gamma)^2 \right>
\propto t^{0.82}$.
Such a sub-diffusion behavior is typically seen in
transient phenomena before the attainment of the steady state.
The interpretation of this anomalous diffusion for the fast mode is difficult.
The sub-diffusion may be due to the combination of the gyro and TTD resonances,
because we observe normal diffusion behavior in cases without the gyroresonance
(section \ref{large}).
With $\xi \simeq (\mathcal{V}_{\rm A}/\overline{V})^2=100$,
the TTD timescale is roughly written with
$t_{\rm TTD} \sim \xi \gamma mc/(e B_0)=10^3 mc/(e B_0)$
in our parameter set. Actually, for $t<t_{\rm TTD}$,
the super-diffusion behavior
($\left< (\Delta \gamma)^2 \right>$ evolves faster than $\propto t$)
is observed in our fast-mode case.
In this relatively long timescale, $t_{\rm TTD}$,
the pitch angle evolution by the gyroresonance
may be not diffusive under the influence of the mirror force.
In spite of this anomalous behavior,
we fit the numerical results with $\left< (\Delta \gamma)^2 \right> \propto t$
as Figure \ref{dgg-evol}, and obtain approximate values
of the diffusion coefficient.

\begin{table}[!htb]
 \caption{Diffusion coefficients with gyroresonance}
  \label{table:full-dev}
  \centering
  \begin{tabular}{lllllll}
\hline
    Name & Mode &  $q$ & $D_{\gamma\gamma}/D_{\gamma\gamma,0}$ \\
    \hline \hline
    A0	 & Alfv\'en & $5/3$ (slab) & $1.08 \pm 0.01 $ \\
    A1	 & Alfv\'en & $5/3$ & $0.35 \pm 0.01 $ \\
    A2& Alfv\'en & $3/2$ & $0.41 \pm 0.01 $ \\
    f1	& fast & $5/3$ & $2.18 \pm 0.31 $\\
    f2 & fast & $3/2$ &  $2.16 \pm 0.27$ \\
    s1& slow  & $5/3$  & $0.80 \pm 0.06$ \\
    s2 & slow  & $3/2$   &   $0.68 \pm 0.09$ \\
    \hline
  \end{tabular}
  \tablecomments{The common parameters are $\alpha=10$, $\eta=10^{-2}$,
$k_{\rm max}=100 k_{\rm min}$,
$\mathcal{V}_{\rm A}=10^{-2} c$, and $\overline{V}=10^{-3} c$.
The initial Lorentz factor is 10 ($k_{\rm res}=0.1 k_{\rm max}$).}
\end{table}

Based on the discussion in section \ref{theory}, we normalize
the diffusion coefficient by an analytical estimate,
\begin{eqnarray}
D_{\gamma\gamma,0} 
\equiv \frac{\pi}{12} (1+3 f_{\rm res}) (q-1) \left(\frac{\overline{V}}{c}\right)^2
\eta^{q-1} \gamma^q \frac{e B_0}{m c}.
\end{eqnarray}
The statistical errors in Figure \ref{dgg-evol} are negligible.
The errors for $D_{\gamma\gamma}$ come from
the deviation from the ideal behavior $\left< (\Delta \gamma)^2 \right> \propto t$
(see dashed lines in Figure \ref{dgg-evol}).
The results are tabulated in Table \ref{table:full-dev}.
The run A0 is the slab case, where only parallel waves are injected.
In other cases, the distributions of turbulent waves are isotropic.
The obtained diffusion coefficients are consistent
with the analytical estimate $D_{\gamma\gamma,0}$
within a factor of 2 or 3.
As we have expected in section \ref{theoryenedif},
the energy-diffusion for the fast mode is slightly larger.
The relatively large errors for the fast and the slow modes
come from the modulated evolutions of $\left< (\Delta \gamma)^2 \right>$
shown in Figure \ref{dgg-evol}.

\begin{figure}[!htb]
\centering
\epsscale{1.0}
\plotone{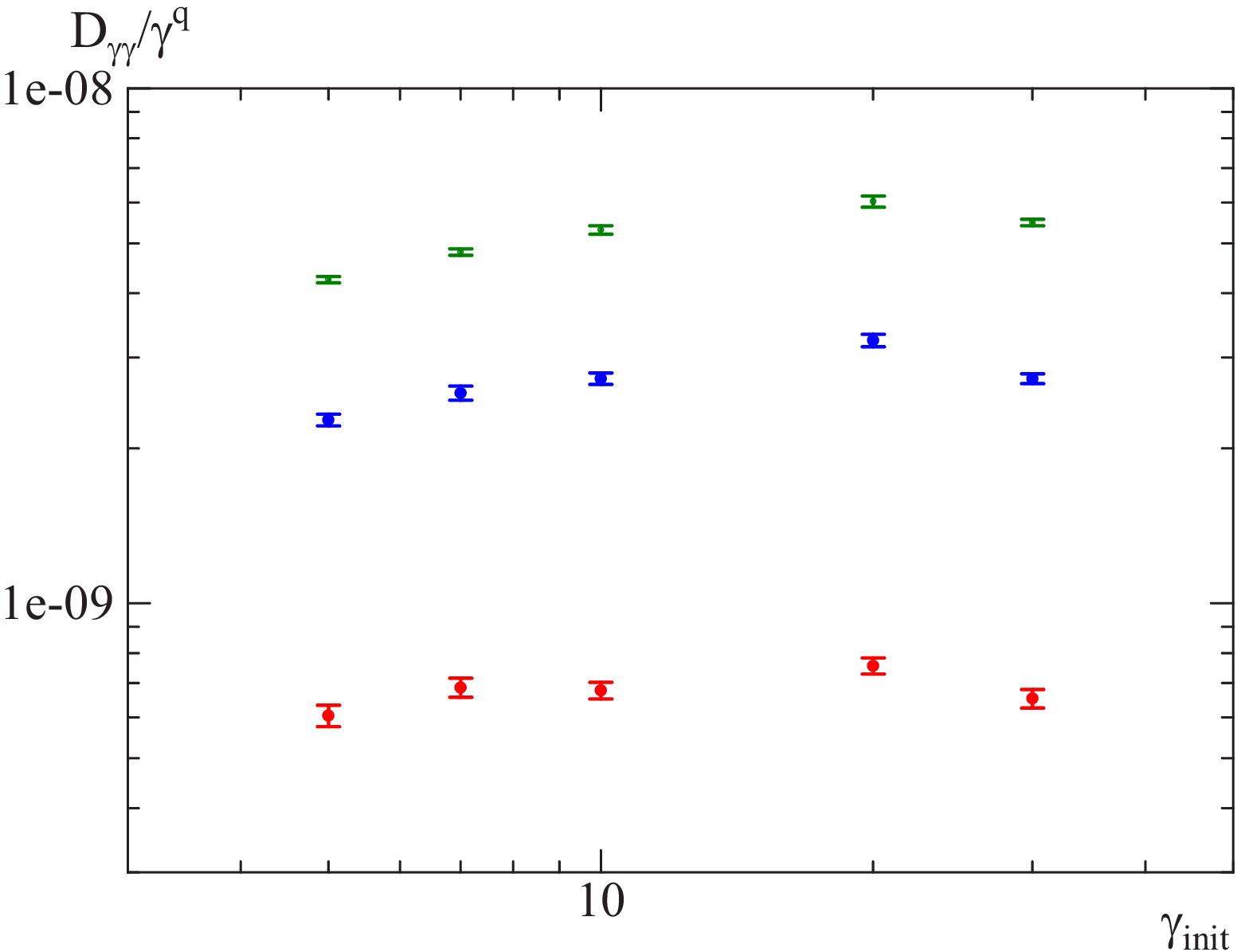}
\plotone{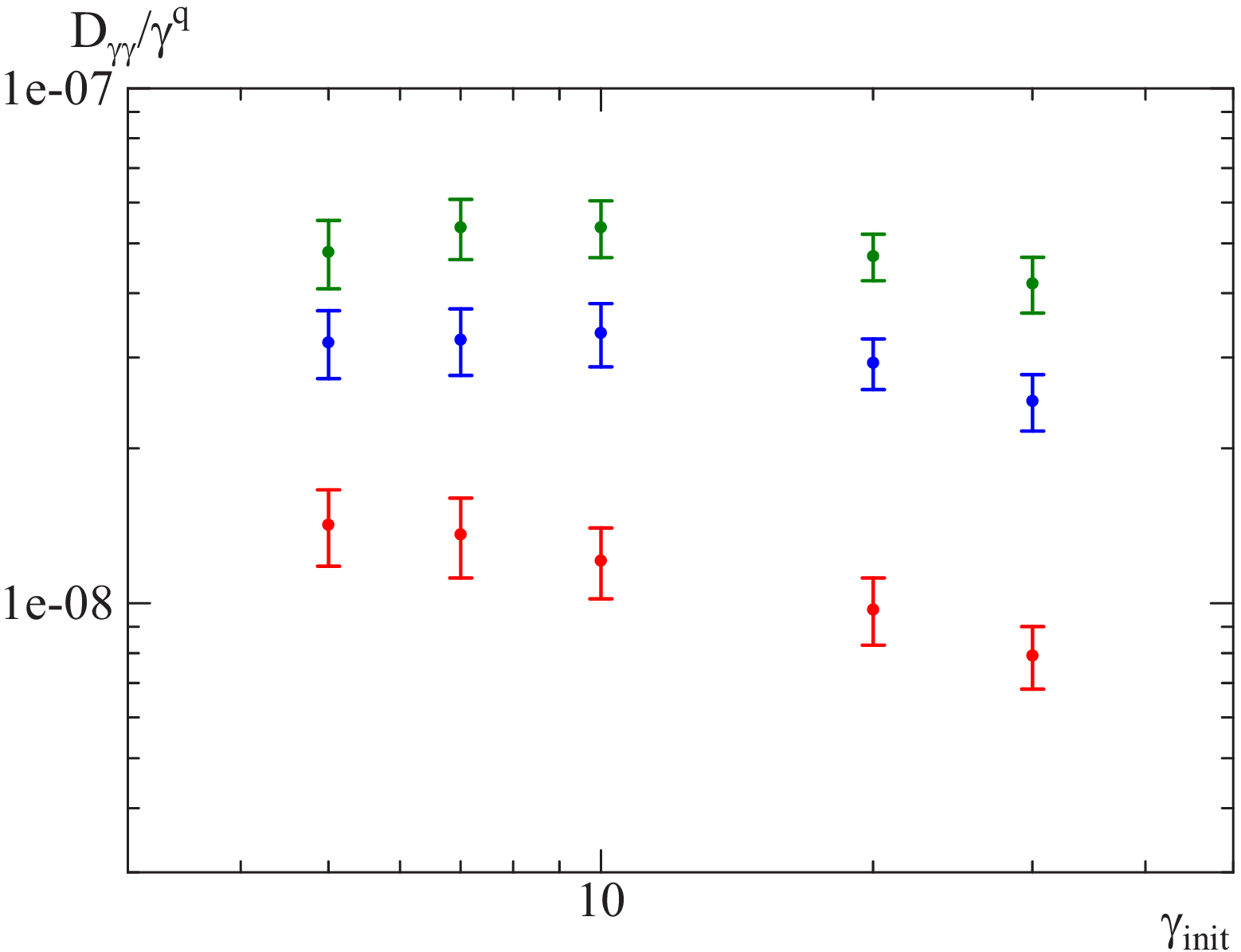}
\plotone{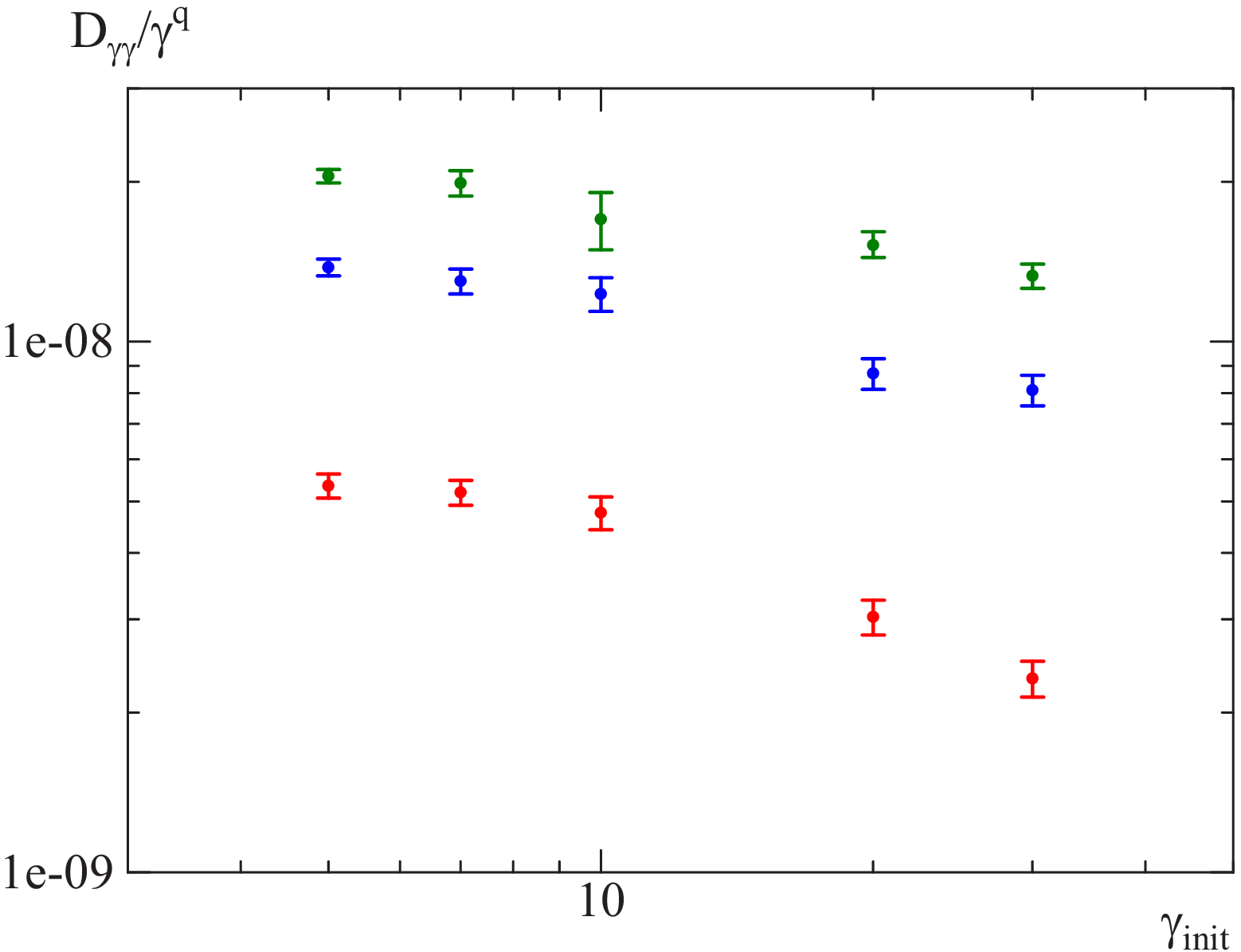}
\caption{Energy dependence of the diffusion coefficient
for the case of $k_{\rm res}<k_{\rm max}$.
The common parameters are $\alpha=10$, $\eta=10^{-2}$,
$k_{\rm max}=100 k_{\rm min}$,
$\mathcal{V}_{\rm A}=10^{-2} c$, and $\overline{V}=10^{-3} c$.
The values are normalized by $e B_0/(mc)$.
The horizontal axis is the initial Lorentz factor.
The red, blue, and green symbols correspond to
$q=2$, $5/3$, and $3/2$, respectively.
The upper, middle, and lower panels show the results for
pure Alfv\'en, fast- and slow-mode waves, respectively.
}
\label{dgg-index}
\end{figure}

To check the energy dependence $D_{\gamma \gamma} \propto \gamma^q$,
we change the initial Lorentz factor within $k_{\rm min}<k_{\rm res}
<k_{\rm max}$.
The results are summarized in Figure \ref{dgg-index}.
Roughly speaking, the results are consistent with the expectation
$D_{\gamma \gamma} \propto \gamma^q$.
For the slow-mode cases and the case $q=2$ in the fast mode,
however, the diffusion coefficient becomes slightly lower
than the anticipated values at a higher $\gamma$.
At present, we do not have clear explanation for those deviations.
As equation (\ref{dgg0}) indicates,
$q=2$ is the threshold value above which
the waves at $k=k_{\rm min}$ dominantly contribute to the TTD process
rather than waves at $k=k_{\rm res}$.
For $q=2$, the lower cutoff in the wavenumber spectrum
may affect the energy dependence of the diffusion coefficient.
In this sense, the results for the slow mode
would suggest that the contribution below $k_{\rm res}$
is larger for the slow mode than the fast mode.
In addition, as we have mentioned,
the coexistence of the gyroresonance and TTD resonance,
whose timescales are different, may cause anomalous effect
on the diffusion.

\subsection{Without Gyroresonance}
\label{large}

In this subsection, we shift the wavenumber distribution
toward a longer wavelength, adopting $\eta=10^{-4}$
and $k_{\rm max}=100 k_{\rm min}$.
For $\gamma=10$, $k_{\rm res}=10 k_{\rm max}$
so that gyroresonance does not occur.
We have confirmed that the energy-diffusion
is not seen in the Alfv\'en mode.
Differently from the cases with the gyroresonance,
the energy dispersion evolution due to fast waves is
almost diffusive ($\left< (\Delta \gamma)^2 \right> \propto t$).
For the fast and the slow modes,
our results are summarized
in Table \ref{table:trans-Afs}, where we normalize the diffusion coefficient by
\begin{equation}
D_{\gamma\gamma,0} \equiv \frac{\pi}{4}  f_{\rm res} (q-1) \gamma^2 \left(\frac{\overline{V}}{c}\right)^2 c k_{\rm max}  
 	\left(\frac{k_{\rm max}}{k_{\rm min}}\right)^{1-q},
\label{eqhs}
\end{equation}
obtained from the discussion in section \ref{theoryenedif}.
While the diffusion coefficients for the fast mode
are slightly larger than the analytical estimate $D_{\gamma\gamma,0}$,
the results for the slow mode well agree with that.
The enhancement of $D_{\gamma\gamma}$ in the fast mode
may be due to the relatively larger electric field induced by waves
with a large $k_\perp$ as mentioned in section \ref{theoryenedif}.

\begin{table}[hbtp]
 \caption{Diffusion coefficients without gyroresonance}
  \label{table:trans-Afs}
  \centering
  \begin{tabular}{lllll}
\hline
    Name & Mode &  $q$ & $D_{\gamma\gamma}/D_{\gamma\gamma,0}$ \\
    \hline \hline
    f3 & fast  & $5/3$   &    $5.59 \pm 0.41$ \\
    f4 & fast  &  $3/2$  &  $4.46 \pm 0.14$ \\
    s3 & slow  & $5/3$  &  $1.13 \pm 0.11$ \\
    s4 & slow  & $3/2$ &  $0.91 \pm 0.09$ \\
    \hline
  \end{tabular}
  \tablecomments{The common parameters are $\alpha=10$, $\eta=10^{-4}$,
$k_{\rm max}=100 k_{\rm min}$,
$\mathcal{V}_{\rm A}=10^{-2} c$, and $\overline{V}=10^{-3} c$.
The initial Lorentz factor is 10 ($k_{\rm res}=10 k_{\rm max}$).}
\end{table}

\begin{figure}[!htb]
\centering
\epsscale{1.0}
\plotone{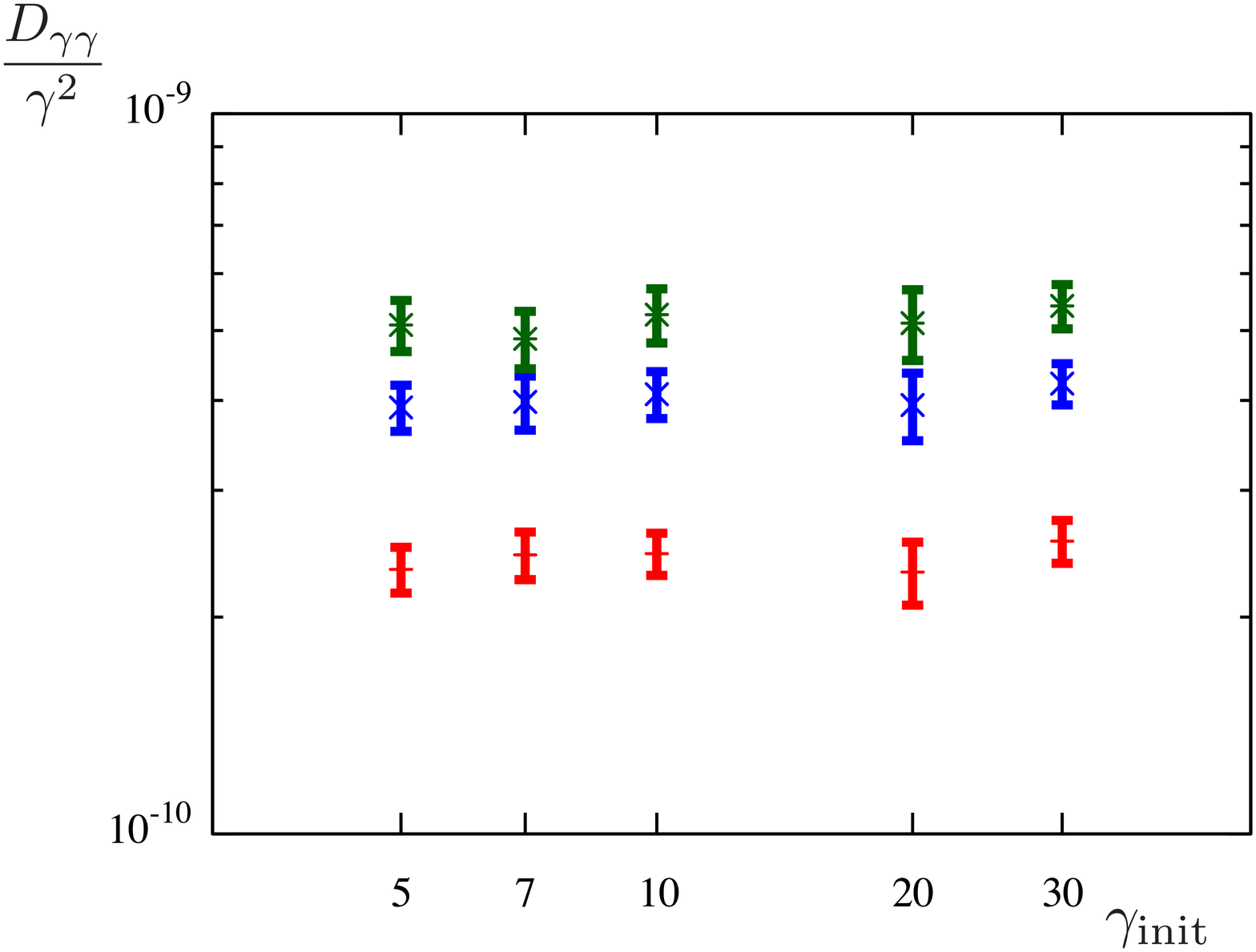}
\caption{Energy dependence of the diffusion coefficient
for the fast-mode cases with $k_{\rm res}>k_{\rm max}$.
The common parameters are $\alpha=10$, $\eta=10^{-4}$,
$k_{\rm max}=100 k_{\rm min}$,
$\mathcal{V}_{\rm A}=10^{-2} c$, and $\overline{V}=10^{-3} c$.
The horizontal axis is the initial Lorentz factor.
The red, blue, and green symbols correspond to
$q=2$, $5/3$, and $3/2$, respectively.
}
\label{dgg-index-large}
\end{figure}

\begin{figure*}[!htb]
\centering
\epsscale{1.0}
\plottwo{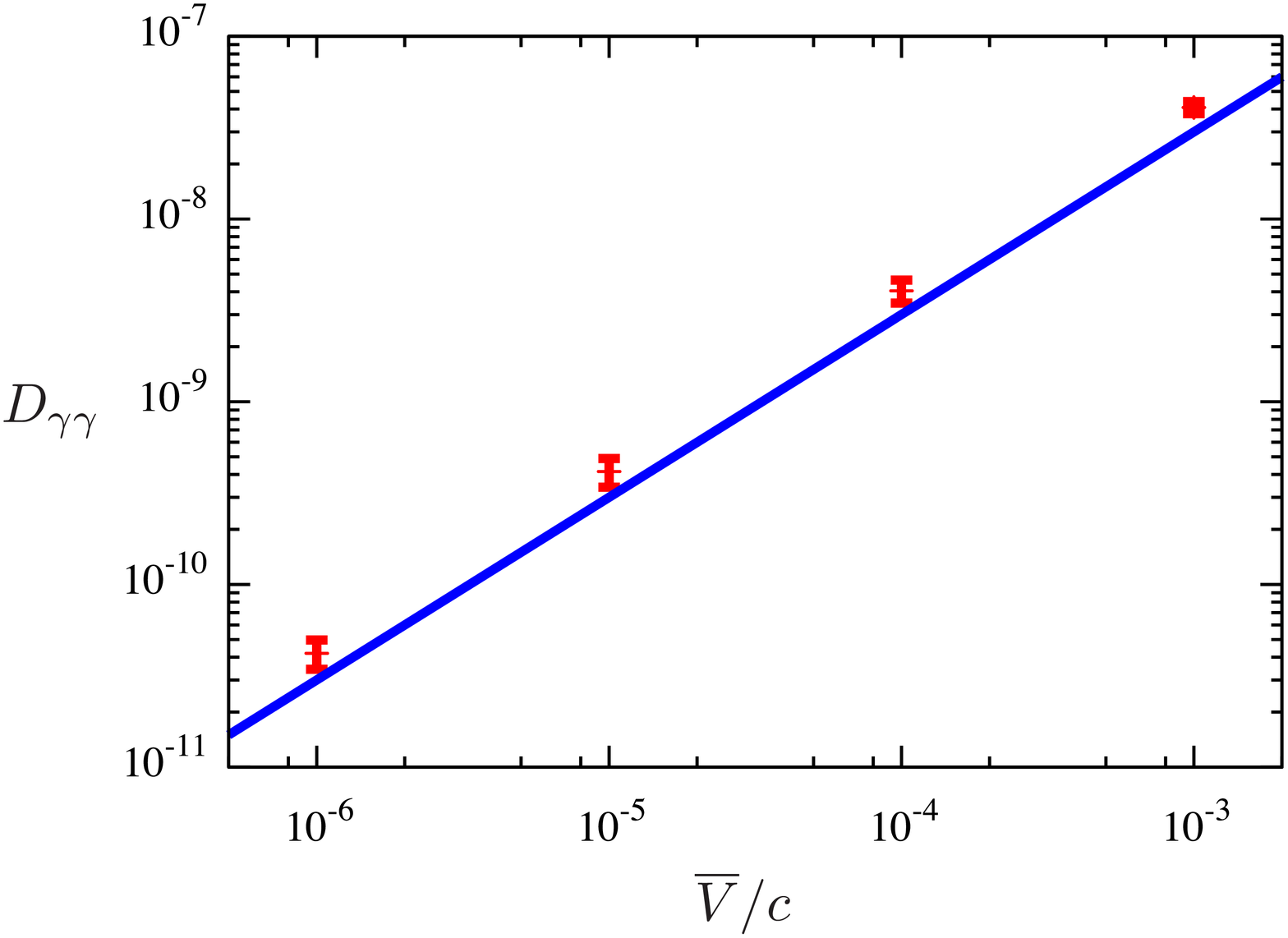}{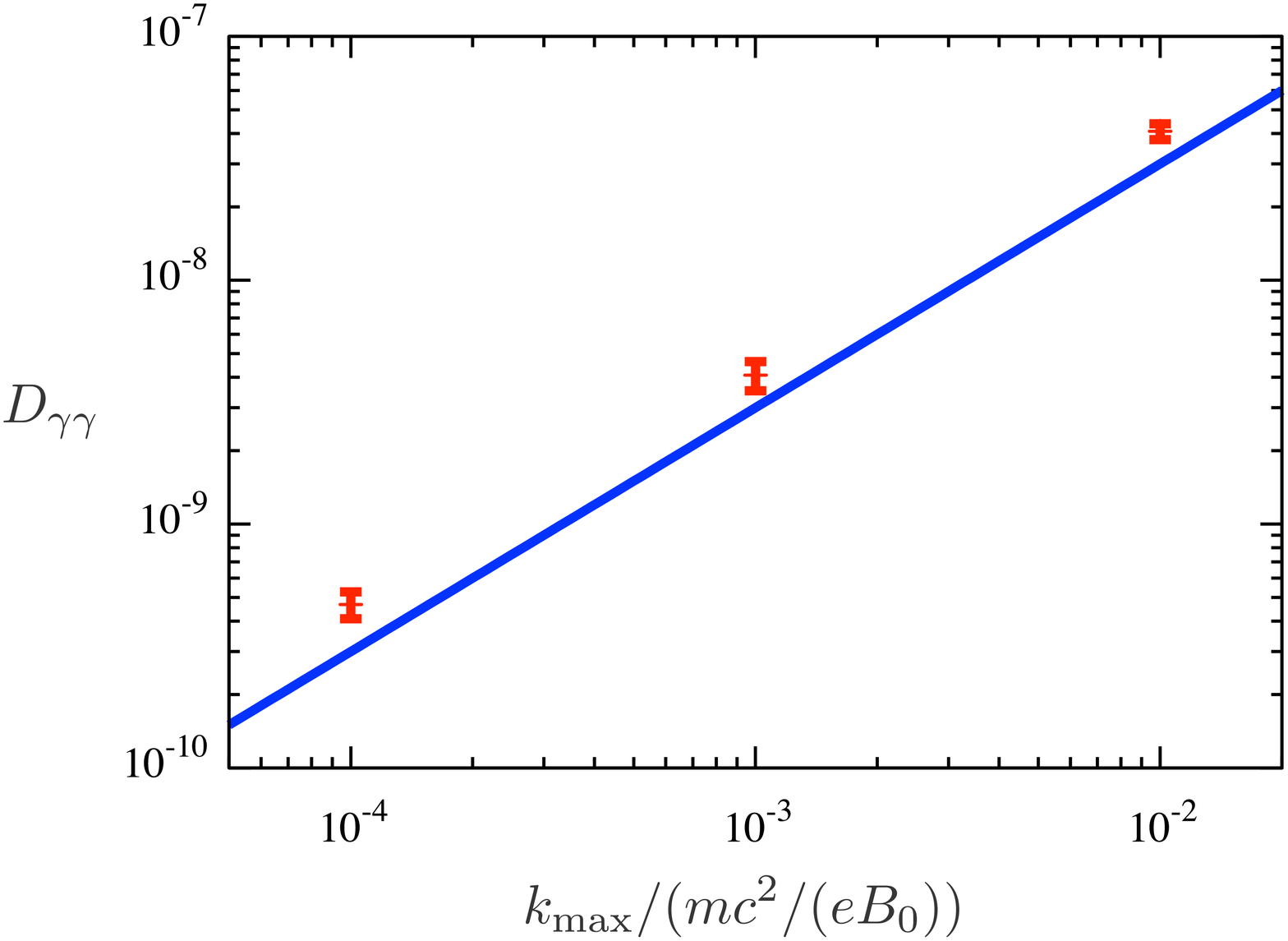}
\plottwo{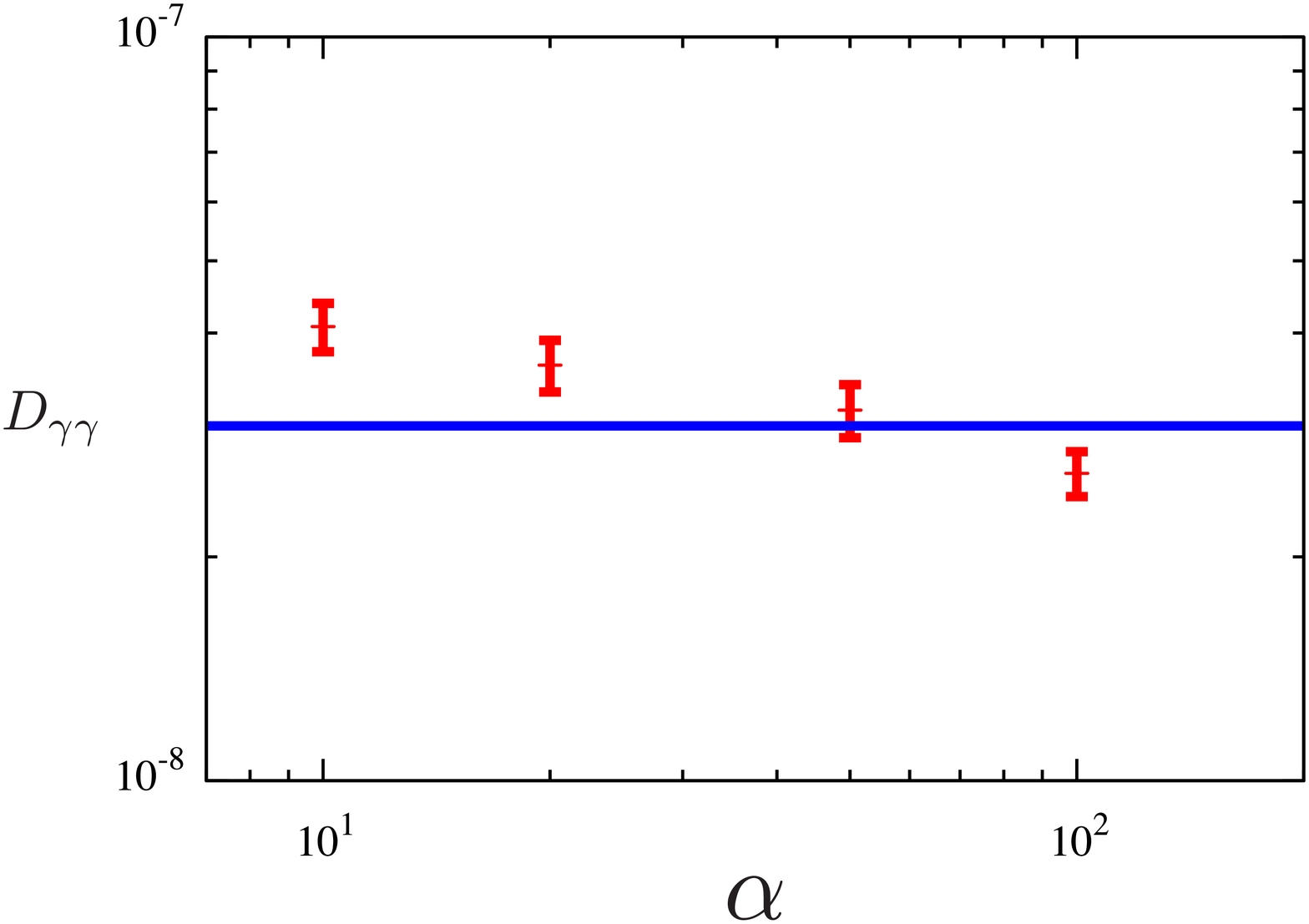}{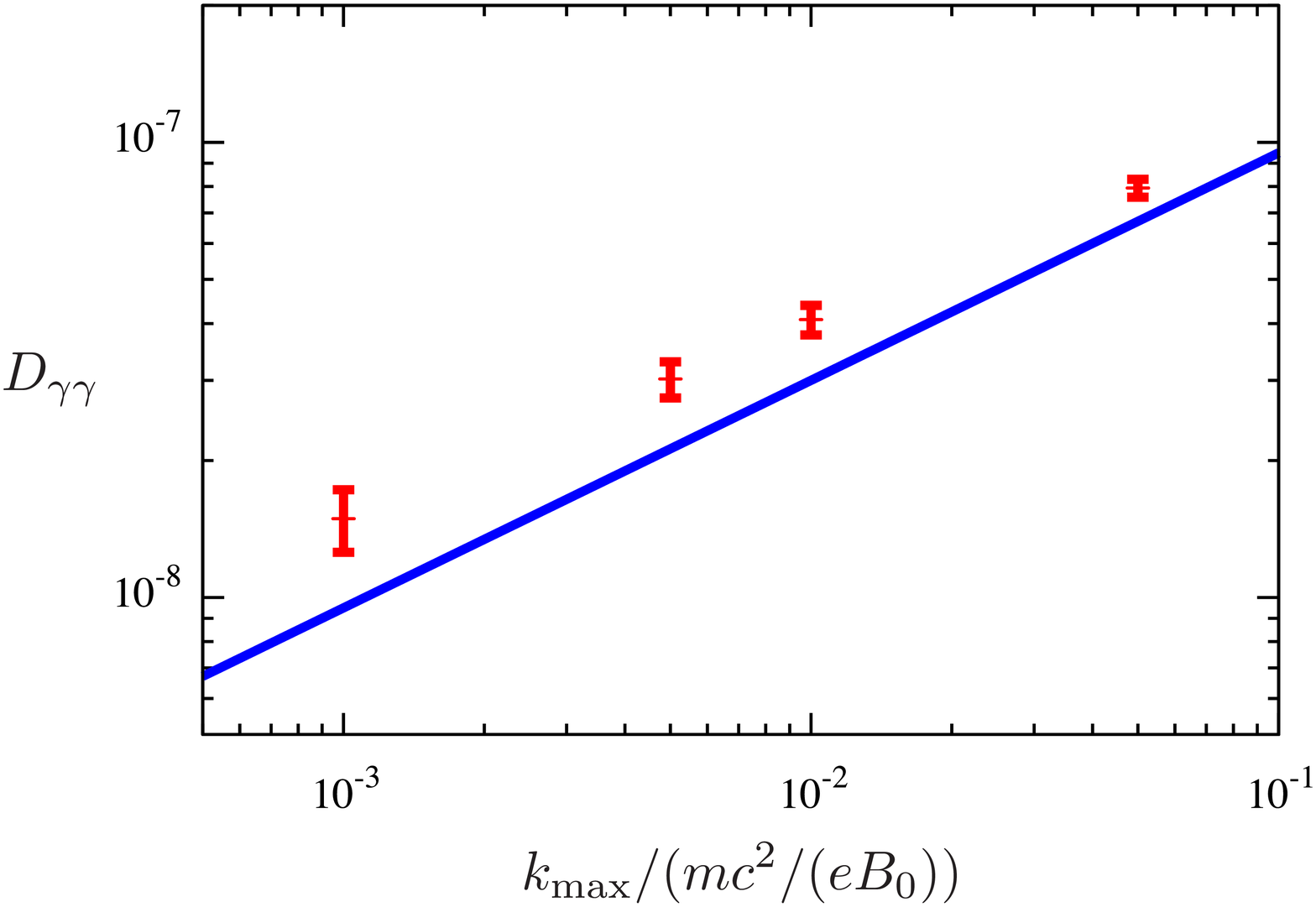}
\caption{Parameter dependence on the diffusion coefficient
at $\gamma=10$
for the fast-mode cases with $k_{\rm res}>k_{\rm max}$.
The fiducial parameters are $q=5/3$, $\alpha=10$, $\eta=10^{-4}$,
$k_{\rm max}=100 k_{\rm min}$,
$\mathcal{V}_{\rm A}=10^{-2} c$, and $\overline{V}=10^{-3} c$.
The plotted values are obtained by changing one of those parameters:
(Upper left)~ $\overline{V}$-dependence. The solid line shows $D_{\gamma \gamma} \propto \overline{V}^2$.
(Upper right)~ $k_{\rm max}$-dependence with a fixed $k_{\rm max}/k_{\rm min}=100$.
The solid line shows $D_{\gamma \gamma} \propto k_{\rm max}$.
(Lower left)~ $\alpha$-dependence. The solid line shows $D_{\gamma \gamma} \propto \alpha^0$.
(Lower right)~ $k_{\rm max}$-dependence with a fixed $k_{\rm min}$.
The solid line shows $D_{\gamma \gamma} \propto k_{\rm max}^{1/2}$.
}
\label{dgg-para-large}
\end{figure*}

The energy dependence of the diffusion coefficients
shown in Figure \ref{dgg-index-large} is consistent with
the hard-sphere-like acceleration irrespectively of the index $q$.
The dependences on the other parameters are also shown in
Figure \ref{dgg-para-large}.
In the analytical estimate in equation (\ref{eqhs}), the diffusion coefficient does not
depend on $\alpha$, but our results suggest
a weak dependence on $\alpha$.
For the other parameters,
Figure \ref{dgg-para-large} roughly supports the analytical estimate.

\section{Application of the TTD-only Model}
\label{application}
In this paper, we have quantitatively estimated the energy-diffusion
of particles via interaction with MHD waves.
The combination of the effects of the acceleration
and the escape from the acceleration site can produce
various type of particle energy distribution
\citep[e.g.][]{byk96,bec06}.
While the gyroresonance with Alfv\'en waves has
been frequently considered as the dominant contribution
to the particle acceleration in high-energy objects
\citep[e.g.][]{sp08,lef11,kak15},
the magnetic energy in such objects is usually subdominant
so that the energy of Alfv\'en waves seems not enough to
account for the energy source of high-energy particles.
Our simulations have demonstrated significant efficiency
of the energy-diffusion
by fast/slow waves without gyroresonance,
which is not widely recognized.
Such non-gyroresonance (TTD-only) interactions with compressible waves
bring us an alternative picture for particle acceleration
in high-energy objects.

The simple hard-sphere acceleration model without
escape effect in Asano \& Hayashida \citep[2018; see also][]{asa15}
can reproduce various blazar spectra.
As an example, let us discuss the model parameters for the representative blazar
Mrk 421 in terms of the particle energy-diffusion
by compressible waves without gyroresonance.
In the model of \citet{asa18}, in which the broadband spectrum
is reproduced with the hard-sphere acceleration,
the comoving size of the acceleration/emission region $R$,
the magnetic field, and the acceleration timescale $\gamma^2/D_{\gamma \gamma}$
are $10^{16}~\mbox{cm}$, $0.18~\mbox{G}$,
and $2.1 \times 10^5~\mbox{s}$, respectively.
The magnetic energy density is much smaller than
the electron energy density so that the turbulence energy
may be dominated by fast waves.

Supersonic turbulences promptly decay via shock waves.
In the relativistic outflow of Mrk 421, therefore,
we can expect that the turbulence velocity is slower than
the sound velocity as $\overline{V}< c/\sqrt{3}$.
Adopting $\xi=1$ and $k_{\rm min}=R^{-1}$,
equation (\ref{eqhs}) and Table \ref{table:trans-Afs} give us
\begin{equation}
\frac{\gamma^2}{D_{\gamma\gamma}}
\simeq 2.1 \times 10^5 \left( \frac{\overline{V}}
{0.2 c} \right)^{-2}
\left( \frac{k_{\rm max}^{-1}}
{1.4 \times 10^{12}~\mbox{cm}} \right)^{1/3}~\mbox{s},
\end{equation}
for $q=5/3$.
The shortest wavelength $k_{\rm max}^{-1}=1.4 \times 10^{12}~\mbox{cm}$
corresponds to the Larmor radius of 76 TeV electron/proton,
while the maximum energy of electrons in Mrk 421
is less than TeV.
With a significantly large cutoff scale of the turbulence
($\gtrsim 10^{12}~\mbox{cm}$) and slower turbulence velocity
than the sound speed,
the particle acceleration required in \citet{asa18} can be achieved.
The cutoff scale $k_{\rm max}^{-1}$ may corresponds to the scale where the
cascade timescale is comparable to the energy transfer timescale
to non-thermal or thermal particles.

\citet{tan17} also applied the SA models
to the radio emission in the Crab nebula.
In that model, parameters are time-dependent,
and electrons are accelerated
during initial a few hundreds years.
The nebula expansion speed is $1800~\mbox{km}~\mbox{s}^{-1}$.
The obtained electron spectrum is consistent with the hard radio spectrum.
The required acceleration timescale \citep[Model 2 in][]{tan17}
is $\gamma^2/D_{\gamma \gamma} \simeq 140~\mbox{yr}$ at the age of 100 yr,
at which the size of the nebula and the magnetic field
are $R_{\rm N}=5.7\times 10^{17}$ cm and $B=1.4$ mG, respectively.
Assuming $q=5/3$, $\xi = 1$ and $k_{\rm min}^{-1}=R_{\rm N}/10$
with equation (\ref{eqhs}) and Table \ref{table:trans-Afs},
we consistently obtain
\begin{equation}
\frac{\gamma^2}{D_{\gamma\gamma}}
\simeq 140 \left( \frac{\overline{V}}
{1800~\mbox{km}~\mbox{s}^{-1}} \right)^{-2} \left( \frac{k_{\rm max}^{-1}}
{3 \times 10^{14}~\mbox{cm}} \right)^{1/3}~\mbox{yr}.
\end{equation}
The required cutoff scale of $3 \times 10^{14}~\mbox{cm}$
is larger than the Larmor radius of electrons whose energy
is less than $130$ TeV.
Therefore, our estimate only with the fast-wave TTD resonance
is consistent for radio-emitting particles,
whose energy are typically lower than 10 GeV.

\section{Discussion}
\label{discuss}

In this paper,
we have quantitatively estimated the energy-diffusion of particles
in temporally evolving pure Alfv\'en, fast- or slow-mode waves.
Particles can interact with waves via not only the gyroresonance but also the TTD resonance.
We have confirmed the significant TTD resonance broadening due to
the finite wave amplitude.
If the particle Larmor radius is large enough for gyroresonance,
both the gyroresonance and the TTD resonance can significantly contribute
to changing the particle energy.
In this case, while the coexistence of the gyroresonance and TTD resonance
leads to a non-standard energy-diffusion for the fast mode,
the slow-mode case is roughly consistent with the diffusion picture.

In the gyroresonance case,
where the turbulence cascade develops to a smaller scale than the Larmor radius,
the dominant contribution to the particle acceleration comes from 
waves with a scale $\sim r_{\rm L}$.
Although a larger scale component can affect the particle energy with the same level,
the trajectory of particles is strongly affected by smaller components than
the larger scale.
While we have assumed that the interaction time $\delta t $ is $1/k_\parallel v_\parallel$, 
it would be shortened by components with a shorter wavelength.
As a result, waves with a wavelength not so far from $r_{\rm L}$
are regarded as the dominant contribution to diffuse the particle energy.
Note that waves with perpendicular scales smaller than $r_{\rm L}$ do not accelerate 
particles efficiently \citep{ch00}.

When the Larmor radius is shorter than the shortest wavelength,
only the TTD resonance contributes to the energy-diffusion.
We have demonstrated that the dominant contribution comes from the smallest scale in
turbulence at $k_{\rm max}^{-1} > r_{\rm L}$ ($D_{\gamma\gamma} \propto k_{\rm max}^{2-q}$).
Our simulations show a slightly larger energy-diffusion
for the fast mode than the simple analytical estimate.
In some high-energy objects, such as blazars, pulsar wind nebulae,
and gamma-ray bursts, particle acceleration due to the TTD resonance
can be expected.
With $\delta B \leq B_0$,
the energy density of Alfv\'en and slow-mode waves is comparable to or less than
the magnetic energy density. In high-energy objects such as blazars,
the magnetic energy density is estimated to be much less than the energy of
the accelerated particles. In such cases, fast-mode waves
are most likely the energy source of high-energy particles.
The TTD-only resonance leads to the hard-sphere-like energy-diffusion coefficient,
which seems consistent with blazar spectra as shown in \citet{asa18}.

To extract the essential physics, we have concentrated on the simplest cases in this paper;
we have neglected the effects of wave damping, anisotropic wave distribution,
and nonlinearity.
\citet{lq14} claimed that all scales equally contribute to accelerating particles
($D_{\gamma\gamma} \propto \ln(k_{\rm max})$)
in slow waves with the resonance broadening due to wave damping.
We need to distinguish the broadening mechanism in MHD simulations,
which has not yet been clarified.
At large scales, turbulence may be hydrodynamical eddy-turbulence
rather than MHD-like wave-turbulence.
In such cases, the particle acceleration mechanism by the fluid compression
may be more relevant as discussed in \citet{pt88} and \citet{cl06}.
The combination of the above multiple processes makes it difficult
to extract the essential acceleration mechanism.
The idealized simulations in this paper are the first step
to find the dominant mechanism of the particle acceleration
in turbulence.

As numerical simulations in \citet{cl03} have shown
\citep[see also][]{clv02,kl10}, while fast-mode waves
distribute isotropically, the slow and Alfv\'enic turbulence is likely anisotropic,
which is typically expressed by the GS model \citep{gs95}.
In the GS turbulence, the perpendicular waves dominate
as $k_\parallel \propto k_\perp^{2/3}$.
At smaller scales than the injection scale, $k_\perp \gg k_\parallel$,
in other words, eddies in small scales are elongated along the background magnetic field.
Although the power-law indices of turbulence are under the discussion \citep{bs05,bs06,be14}, 
the anisotropy is confirmed by many direct MHD simulations
\citep[for a review,][]{bl13}.

\citet{ch00} has shown that the acceleration efficiency  
in the GS turbulence is much lower than that for isotropic turbulence.
Even when the gyroresonance $k_\parallel^{-1} \sim r_{\rm L}$ is realized,
the short perpendicular scale
leads low efficiency of acceleration
because of too frequent changes of the electric field in one gyro motion.
On the other hand, the TTD resonance condition is broadened by the mirror force non-resonantly,
and the GS anisotropy makes the maximum energy-absorption condition of $r_{\rm L} \sim k_\perp^{-1}$
at a large scale $k^{-1} \sim k_\parallel^{-1} \gg k_\perp^{-1}$, which enhances
the energy-diffusion efficiency.
\citet{ber11} estimated the spatial diffusion of particles 
following test particles in turbulences generated from direct incompressive MHD simulations.
They conclude that the main scattering process is the TTD with the pseudo-Alfv\'en mode,
which is the incompressive limit of the slow mode,
and the gyroresonance is inefficient in the GS turbulences.

In the GS turbulence, as discussed above, 
the TTD resonance with slow-mode waves rather than the gyroresonance with Alfv\'en waves
dominantly contributes to the particle energy-diffusion.
Considering the negative effect of the anisotropy in slow and Alfv\'en waves,
here we again emphasize the importance of the fast mode in the particle energy-diffusion.

\section*{Acknowledgment}
First, we appreciate the two anonymous referees for their
helpful advice.
Numerical computation in this work was carried out at the Yukawa Institute Computer Facility.
This work is supported by JSPS KAKENHI grant No.
16K05291, and 18K03665 (KA). This work is carried out by the joint research program
of the Institute for Cosmic Ray Research (ICRR),
the University of Tokyo.

\end{document}